\documentclass[5p, sort&compress]{elsarticle}

\makeatletter
\def\@author#1{\g@addto@macro\elsauthors{\normalsize%
	\def\baselinestretch{1}%
    \upshape\authorsep#1\unskip\textsuperscript{%
      \ifx\@fnmark\@empty\else\unskip\sep\@fnmark\let\sep=,\fi
      \ifx\@corref\@empty\else\unskip\sep\@corref\let\sep=,\fi
      }%
    \def\authorsep{\unskip,\space}%
    \global\let\@fnmark\@empty
    \global\let\@corref\@empty  %% Added
    \global\let\sep\@empty}%
    \@eadauthor={#1}
}
    
\def\ps@pprintTitle{%
 \let\@oddhead\@empty
 \let\@evenhead\@empty
 \def\@oddfoot{}%
 \let\@evenfoot\@oddfoot}
\makeatother

\usepackage{placeins}
\usepackage{lineno,hyperref,graphicx,color,amssymb,wasysym}
\usepackage[flushleft]{threeparttable}

\modulolinenumbers[5]

\journal{Astroparticle Physics}

\begin{document}\sloppy
%issues with line breaking unless I add sloppy which relaxes the rules for inter word spacing

\begin{frontmatter}

\title{Study and mitigation of spurious electron emission \\ 
from cathodic wires in noble liquid time projection chambers}

\author{A. Tom{\'a}s\corref{correspondingauthor}}
\cortext[correspondingauthor]{Corresponding author}
\ead{a.tomas@imperial.ac.uk}
\author{H. M. Ara{\'u}jo}
\author{A. J. Bailey}
\author{A. Bayer}
\author{E. Chen}
\author{B. L{\'o}pez Paredes}
\author{T. J. Sumner}
\address{High Energy Physics Group, Blackett Laboratory, Imperial College London, London SW7 2AZ, U.K.}

%\date{\today}

\begin{abstract}
Noble liquid radiation detectors have long been afflicted by spurious electron emission from their cathodic electrodes. This phenomenon must be understood and mitigated in the next generation of liquid xenon (LXe) experiments searching for WIMP dark matter or neutrinoless double beta decay, and in the large liquid argon (LAr) detectors for the long-baseline neutrino programmes. We present a systematic study of this spurious emission involving a series of slow voltage-ramping tests on fine metal wires immersed in a two-phase xenon time projection chamber with single electron sensitivity. Emission currents as low as $10^{-18}$~A can thus be detected by electron counting, a vast improvement over previous dedicated measurements. Emission episodes were recorded at surface fields as low as $\sim$10~kV/cm in some wires and observed to have complex emission patterns, with average rates of 10--200~counts per second (c/s) and outbreaks as high as $\sim$10$^6$~c/s. A fainter, less variable type of emission was also present in all untreated samples. There is evidence of a partial conditioning effect, with subsequent tests yielding on average fewer emitters occurring at different fields for the same wire. We find no evidence for an intrinsic threshold particular to the metal-LXe interface which might have limited previous experiments up to fields of at least 160~kV/cm. The general phenomenology is not consistent with enhanced field emission from microscopic filaments, but it appears instead to be related to the quality of the wire surface in terms of corrosion and the nature of its oxide layer. This study concludes that some surface treatments, in particular nitric acid cleaning applied to stainless steel wires, can bring about at least order-of-magnitude improvements in overall electron emission rates, and this should help the next generation of detectors achieve the required electrostatic performance.
\end{abstract}


\end{frontmatter}

%\linenumbers

%%%%%%%%%%%%%%%%%%%%%%%%%%%%%%%%%%%%%%%%%%%%%%%%%%%%%%%%%%%%%%%%%%
\section{Introduction}

Two-phase xenon detectors employed in direct dark matter searches, such as that being developed for the LUX-ZEPLIN (LZ) experiment~\cite{LZTDR} which motivated this study, require extremely low thresholds for scintillation and ionisation signals, at the level of single quanta~\cite{AprileDoke10,ChepelAraujo13}. Although this technology has been at the forefront of the field for several years, it remains the case that most such LXe Time Projection Chambers (LXe-TPCs) have not been able to operate at their design electric fields. Similar problems have afflicted this technology applied to neutrinoless double beta decay searches, as well as LAr instruments for both dark matter and neutrino detection. Although these difficulties have long been recognised, a definitive explanation is still lacking. A review of the various high-voltage (HV) issues affecting these communities can be found in Ref.~\cite{Rebel14}.

Aside from any difficulties related to the HV feed\-through technology, where progress has been made in recent years, the prominent limitation which has often materialised can be traced to the poorly understood emission of light and charge specifically from the wire grids used to define the various field regions. In two-phase detectors these include the cathode wire-grid at the bottom of the sensitive volume and the gate wire-grid located just below the liquid surface, which are both cathodic electrodes. They define a `drift field' between them which sweeps the ionisation released by particle interactions in the active liquid volume; the gate also strengthens the `extraction field' below the liquid surface, promoting electron emission and the subsequent generation of the electroluminescence response in the vapour phase. These grids are typically made from parallel or woven metal wires with diameters of tens to hundreds of micrometers. Spurious emissions are observed at electric fields as low as 10~kV/cm on the wire surface~\footnote{Surface fields are calculated for the perfect (cylindrical) wire geometry unless indicated otherwise.}, much lower than expected for phenomena such as liquid-phase electroluminescence and field emission; these have prevented the operation of previous instruments at their design voltages: the threshold of these dark matter detectors is so low (especially in the ionisation channel) that the emission of individual electrons and photons can interfere with the physics searches, which aim to detect $\sim$keV energy deposits containing very few scintillation photons and ionisation electrons. Spurious emissions from electrode grids also jeopardise the so-called `S2-only' searches (using electron counting below the scintillation threshold)~\cite{Hagmann04,Santos11,Essig12}, as in this type of analysis the accurate reconstruction of the vertical coordinate is not possible.

Although detection thresholds for ionisation are typically much higher in the large LAr-TPCs for neutrino detection, electrons emitted by any cathodic surface may be drifted long distances and accumulate on dielectric surfaces several metres away and lead to discharges later on. In this instance it is not trivial to diagnose that this process may be taking place at all.

The electrostatic design methodology adopted by previous experiments was thus found to be compromised. This was based on the onsets for electroluminescence and charge multiplication in LXe at \smash{412$^{+10}_{-133}$}~kV/cm and \smash{725$^{+48}_{-139}$}~kV/cm, respectively~\cite{Aprile14a} (see also \cite{Derenzo74,Masuda79}), while practical LXe-TPC cathodes made from stainless steel wires have been limited to surface fields of 40--65~kV/cm~\cite{Howard04,Lebedenko09,Burenkov09,Akimov12a,Akimov12b,Akerib14}. A detector with gold-plated stainless steel wires could not operate at the design field either~\cite{XENON1Tinstrument}. A chamber with Monel wires, known for its resistance to corrosion, sustained 35~kV/cm on the gate grid~\cite{pixey17}. Notably, a small chamber achieved substantially higher fields of 220~kV/cm on BeCu wires~\cite{Shutt07}. Spurious emission has been reported also in detectors using etched meshes instead of wires~(e.g.~\cite{XENON100instrument}).

This study set out to determine the underlying causes for this phenomenology and to attempt to find suitable mitigation by testing systematically a number of fine cathodic wires a few centimetres long for the emission of quanta of light and/or charge in a small two-phase xenon chamber built for this purpose. This work was conducted within the wider R\&D framework for the LZ experiment, in particular in coordination with colleagues at SLAC where larger electrodes are being tested; these involve $\sim$10~m of wire initially, followed by validation of the full LZ grids utilising $\gtrsim$700~m of wire (see Section~3.10 in \cite{LZTDR} for additional information on this programme).

%%%%%%%%%%%%%%%%%%%%%%%%%%%%%%%%%%%%%%%%%%%%%%%%%%%%%%%%%%%%%%%%%%
\section{Experimental Method}

\subsection{Two-phase xenon chamber}

\begin{figure*}[tbh]
\begin{center}
\includegraphics[width=18.0cm]{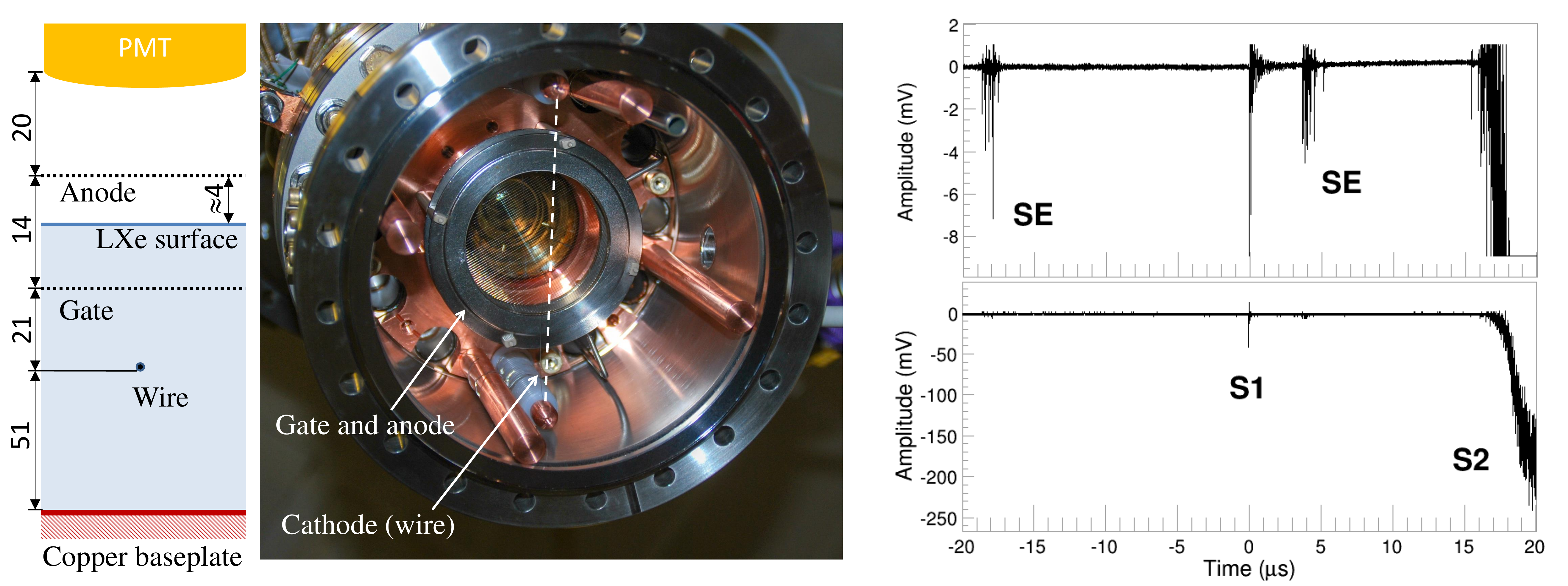}
\end{center}
\caption{Left -- Schematic of the chamber (dimensions in mm). Centre -- Upward view of the chamber with the copper baseplate removed. The PMT is visible through the 60~mm wide gate and anode grids which establish electroluminescence above the liquid surface. The cathode wire sample (highlighted) is stretched between the two feedthroughs as shown. Right -- Typical waveform recorded with a 100-$\mu$m ZEPLIN-III wire with 145~kV/cm on its surface, in the high- and low-gain channels. The ringing visible in the waveform is due to the use of an external PMT voltage divider. A large S2-like signal (seen clearly in the lower, low-gain channel) is preceded by an S1-like optical pulse followed by a SE pulse, attributed to photoionisation by the S1 light. Another SE is registered prior to S1, which constitutes a candidate pulse from spurious emission.}
\label{fig:ChamberPhoto}
\end{figure*}

The small (4~kg) LXe chamber depicted in Fig.~\ref{fig:ChamberPhoto} was developed for these tests. Gate and anode wire-grid electrodes are located just below and above the liquid surface, respectively; these grids were built from 100~$\mu$m wire (SS316L) with 1~mm pitch, oriented at 90$^{\rm o}$ to each other. The cathode wire-grid which would normally exist at the bottom of the active volume was replaced by a single thin wire (the sample under study), located 21~mm below the gate. Thus a strong electric field is achieved at the wire surface with only modest voltages delivered into the chamber. In the basic test, the field on the wire surface is increased steadily, and single electron (SE)-like signals which are unrelated to any particle interaction in the chamber are searched for in the subsequent data analysis. Particle interactions (which are not the focus of this study) are detected via prompt scintillation (S1) and delayed electroluminescence (S2) signatures; an example is shown in Fig.~\ref{fig:ChamberPhoto} (right), along with two SE pulses.

At the bottom of the chamber, a thick copper baseplate is cooled to $\approx\!-\!100^\circ$C by a long `cold finger' immersed into an open LN$_2$ dewar located under the detector, providing 12~W of cooling power. Thermal control is achieved through external heaters attached to the baseplate. Despite that, this is the coldest detector surface, providing a good thermal profile to avoid bubbling. 

A 2-inch, quartz-windowed ETEL~D730/9829Q photomultiplier tube (PMT) is located in the gas phase viewing downward. Electrons released from the upper surface of the test wire are drifted upward past the gate and emitted into the vapour phase, where they generate electroluminescence. This provides sensitivity to single electrons emitted from the wire, which was the key design driver. Although the wire length is 130~mm, the length effectively under test is just over 50~mm (cf.~Section~\ref{subsec:Electrostatics}).

The wire sample is cleaned prior to installation in two consecutive, 1-hour long ultrasonic baths in scientific grade acetone and isopropyl alcohol, then dried with a jet of filtered argon gas. After assembly of the sample, which requires the chamber to be exposed to air for about one hour, the system is warmed to 50$^{\circ}$C and pumped to high vacuum for 2--4 days while monitoring the pressures of electronegative species. Xenon gas is then introduced and circulated through a heated zirconium getter for a further 3~days to ensure sufficient electron lifetime ($\gtrsim$20~$\mu$s). The available cooling power does not permit continuous purification during liquefaction or testing, but the chamber maintains the required electron lifetime without degradation due to the use of only low out-gas materials. Two measurements were done of this lifetime after a LXe run prepared with the above procedure using an external monitor and, in both cases, the value obtained was beyond the dynamic range of the instrument ($\tau\!\approx\!100$~$\mu$s). We were able to determine if a dataset exhibited insufficient purity from data analysis---which occurred only twice, when the above procedure was not adhered to.

The turnaround time between runs was about 2~weeks, with the run itself completed within one day: the chamber is cooled overnight to $-110^\circ$C; liquefaction takes up to 6~hours plus 1~hour to stabilise; emptying the chamber after the testing takes $\approx$1~hour.

\subsection{High voltage delivery}

In most tests the gate was operated at $-$2~kV and the anode at $+$5~kV; the cathode voltage was slowly ramped at a constant 1~V/s from a computer-controlled Bertran 225-50R power supply with a current resolution of 10~nA, set to trip at 30~$\mu$A. All voltages are delivered to the chamber via gas-phase feedthroughs. The two cathode feedthroughs, between which the 130-mm wire sample is stretched as shown in Fig.~\ref{fig:ChamberPhoto} (centre), are Ceramaseal-21130C1-25kV, with an additional 4-mm thick layer of ultra-high molecular weight polyethylene (UHMWPE) with a corrugated lateral surface added around the central post. Each feedthrough is capped by a rounded, 10-mm diameter copper piece with an internal cavity inside which a small screw attaches the wire.

The HV delivery from the gas phase limited the anode and cathode operational voltages, and thus the maximum fields on the wire samples. The maximum (negative) voltage held by the cathode in cold xenon gas at 1.7~bar decreased progressively from 9.5~kV during the first few runs and stabilised in the range 5.1--5.7~kV. An increase in working pressure to 2.45~bar improved this maximum voltage to $\approx$6.3~kV in the last three runs. Light emission from the cathode feedthroughs was occasionally observed immediately prior to the power supply trip at the end of the voltage range. This light was easily distinguishable from electron emission from the wire in data analysis.

\subsection{Optical response} \label{subsec:Optics}

\begin{figure*}[tbh]
\begin{center}
\includegraphics[width=18.0cm]{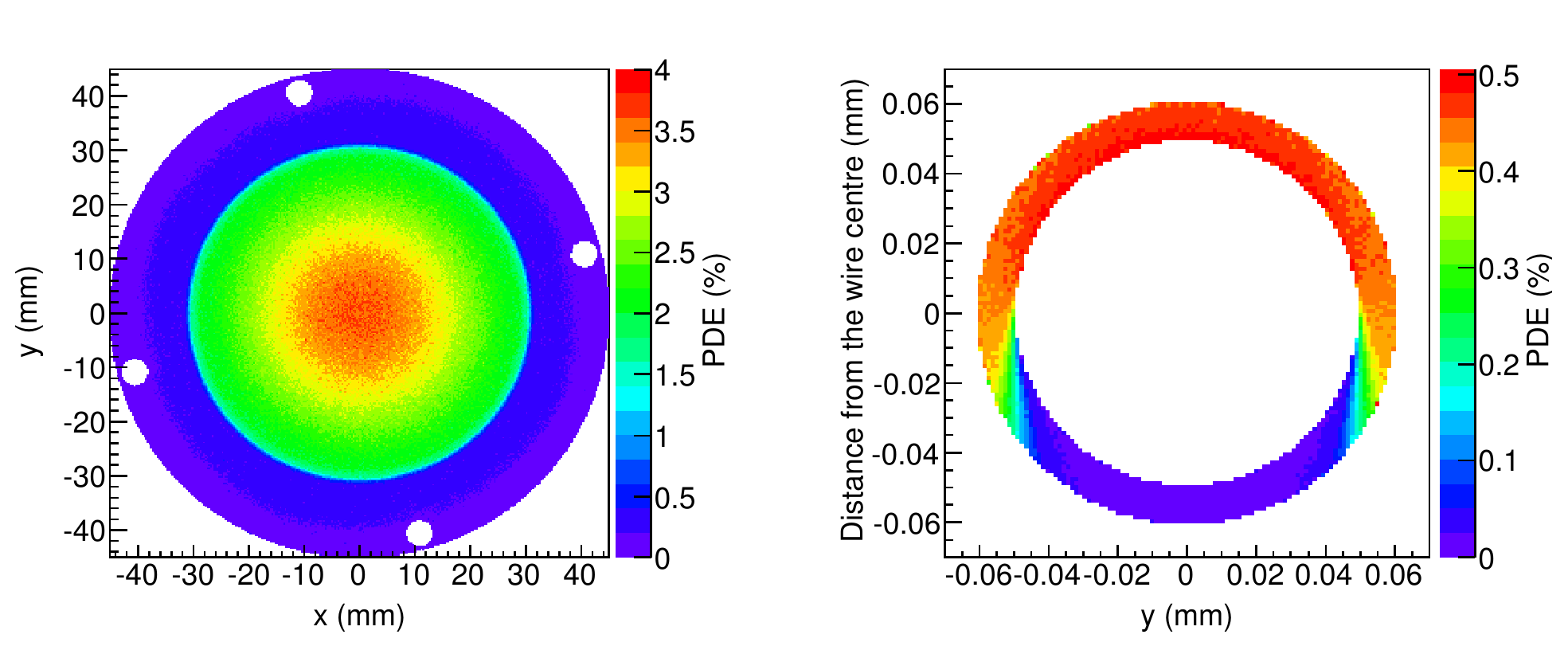}
\end{center}
\caption{Left -- PDE for S2 and SE electroluminescence photons in the horizontal plane; there is a drop of $\approx$2$\times$ between the central region and the edge of the active region. Right -- PDE averaged along the central 52~mm for photons emitted near a 100~$\mu$m cathode wire with 29\% reflectivity; photon emission from the upper side of the wire has $\approx$0.5\% PDE.}
\label{fig:GEANT4}
\end{figure*}

The liquid level was located approximately 10~mm above the gate and 4~mm below the anode. This gap generates $\simeq$280~electroluminescence photons per electron emitted from the liquid at the 1.7~bar operating pressure~\cite{Fonseca04,Boulton17}. Figure~\ref{fig:GEANT4} shows the photon detection efficiency (PDE) for electroluminescence signals (S2 and SE) simulated with GEANT4~\cite{Agostinelli03}; this includes a 30\% PMT quantum efficiency (QE) measured at low temperature~\cite{Araujo04,LopezParedes18}. This PDE varies $<$10\% with the VUV reflectivity chosen for the various materials, given that no high-reflectivity surfaces were installed in the chamber. These simulations predict a marked dependence of the mean SE response on the radial position, with an average of $\simeq$8 photons detected (phd) per electron for uniformly distributed locations inside the 30~mm radius of the gate and anode grids. The SE response distributions obtained experimentally from calibrations in each run varied in the range 8.3--11.5~phd, in reasonable agreement with our prediction.

For SE distributions with means in this range, our analysis threshold at 4~photoelectrons (phe) per pulse represents an SE detection efficiency of $>$85\%. Besides this condition, the pulse size is not used to define the selection criteria, which rely on pulse timing alone. In the later tests at 2.45~bar a wider gas gap $\gtrsim5$~mm compensated for the lower electroluminescence yield. The probability for cross-phase electron emission (often referred to as `extraction efficiency') varied in the range 60--70\% for the gaps used in our tests~\cite{Edwards17}. Overall, the estimated efficiency for SE detection is in the range 50--60\%. The final test in each run was usually conducted with the gate at ground in order to increase the maximum field on the surface of the wire sample; in this instance the SE detection efficiency is only 25--30\%, mainly due to the lower emission probability. Electron loss due to the finite electron lifetime is negligible at our levels of purity.

The chamber is also sensitive to prompt LXe scintillation (S1 pulses) and other optical signals, although the PDE is poor in this instance as the PMT is not well matched optically to the liquid. The PDE for photons emitted near the wire surface is shown in Fig.~\ref{fig:GEANT4}; here we assume a conservative VUV reflectivity of 0.29 for the wire, which is half of the value reported for stainless steel~\cite{Bricola07}. The maximum PDE is only $\simeq$0.5\%, it remains rather constant in the upper half of the wire (facing the PMT) and decreases sharply for its lower surface and more slowly away from the centre. The wire length tested for both light and charge emission are in fact similar.

\subsection{Electrostatics} \label{subsec:Electrostatics}

\begin{figure*}[tbh]
\begin{center}
\includegraphics[width=18.0cm]{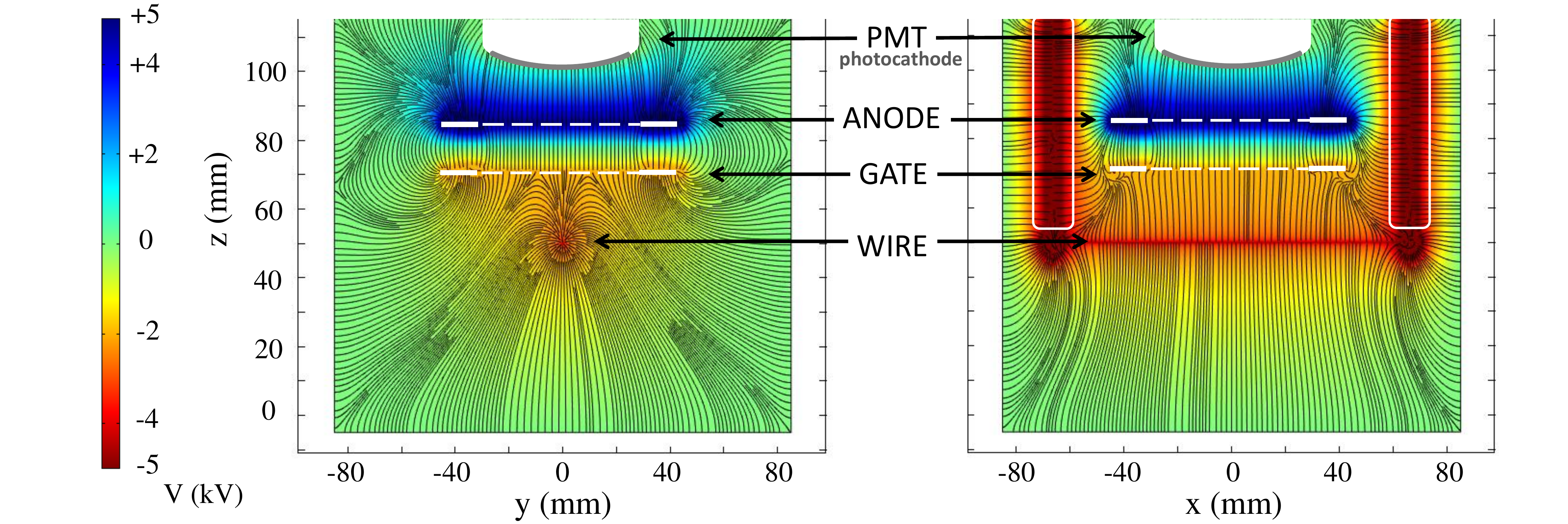}
\end{center}
\caption{Electrostatic model of the chamber obtained with COMSOL. Spatial maps of voltage and charge trajectory lines are shown on transversal cuts across the chamber (perpendicular to the wire on the left and along the wire on the right) for the applied voltages given in the text. The PMT is indicated in white (with the photocathode highlighted in grey) as is the UHMWPE around the cathode feedthroughs and the electroluminescence electrodes: the gate and anode grids (dashed) and their supporting rings (solid). Most trajectories emerging from the top of the wire connect to the gate-anode region almost up to the edge of the grid rings.}
\label{fig:COMSOL}
\end{figure*}

The electric field distribution in the chamber was calculated using COMSOL~\cite{COMSOL} and is shown in Fig.~\ref{fig:COMSOL} for typical test parameters: gate at $-2$~kV, anode at $+5$~kV, and a reasonably high voltage of $-5$~kV on the cathode wire. The electron trajectories in the gas are mostly straight, ensuring a constant electroluminescence gain. Analysis of those electron trajectories emerging from the wire allows us to determine that $\simeq$7\% of its surface is under test---with $\simeq$50\% detection efficiency (c.f. Section~\ref{subsec:Optics})---along the central 52~mm of the sample. This calculation shows $<$0.2\% field variation around a 100~$\mu$m wire, and the strength along this `active length' varies by $<$5\% before dropping sharply near the feedthrough posts.

The electric field at the wire surface was estimated independently using a Garfield++~\cite{Garfield++} model for an infinitely long wire between two infinite parallel plates, representing the gate grid and the bottom of the chamber. The two models are in good agreement. The maximum field established on the wire varied between $\sim$400~kV/cm for the smallest diameter tested (40~$\mu$m) and 60~kV/cm for the largest (500~$\mu$m). Most tests were conducted with a 200~$\mu$m wire, reaching $\sim$70~kV/cm at maximum cathode voltage. The COMSOL-calculated field over the active length can differ by as much as $10\%$ from that obtained with the simpler model depending on the particular combination of voltages; the disagreement is typically $\approx$2\% for the first voltage ramping test that provides the most valuable data, and we use the Garfield++ model to calculate the wire fields in presenting our results as this is more practical. Other systematic errors have been considered and are smaller. The total error on the electric field on the wire is $\approx$5\%.

Two observations are pertinent here, advancing the discussion on possible backgrounds. Firstly, the external edge of the anode collects field lines in a region where electroluminescence can occur, although with low PDE and producing pulse shapes that will differ from standard SE signals. In particular, as the feedthrough posts are encased in UHMWPE they are an unlikely source of electrons. Secondly, SE emission from the lateral chamber volumes can be collected through the periphery of the gate grid, but the PDE is particularly low there.

\subsection{Data acquisition and pulse selection}

The DAQ is an 8-bit dual-range system inherited from ZEPLIN-III from which we record one channel digitised at 250~MS/s (4~ns/sample) before and after a 10$\times$ wideband amplifier. During each test the cathode voltage is slowly ramped at 1~V/s while recording the PMT waveforms triggered at 8~Hz by a pulse generator. A 250~$\mu$s waveform is recorded per trigger; the 125~ms period between waveforms is longer than any slow response mechanism which is likely to operate in the chamber. During the test $\approx$0.18~s of live time (varying with the application of various off-line vetoes) are recorded for every 100~V increase, collecting $\approx$6~s in total. Slow control data including xenon pressures, temperatures and electrode voltages are embedded into the DAQ stream. The data are scanned off-line for photon and electron emission using the ZE3RA software~\cite{Neves11}. A waveform showing electron emission both preceding and during a regular event (with S1 and S2 pulses) was shown in Fig.~\ref{fig:ChamberPhoto} (right).

Selected SE-like pulses must not follow an S1 and the entire waveform must not contain an S2, as these optical stimuli can produce SE signals as discussed below. To reduce the probability of the response to a previous particle interaction spilling over into the search waveform, the initial 30~$\mu$s are disregarded in the search for electron emission---this is longer than the maximum ionisation drift time observed in the chamber ($\sim$20~$\mu$s). SE-like pulses fulfilling these criteria are hereafter termed \textit{emission candidates}. The test sensitivity is determined by the acquired live time, the electron collection efficiency from the wire, the emission probability from the liquid surface, the pulse identification efficiency, and the background rate of SE-like pulses unrelated to wire emission.

The leading sources of the small background ($\lesssim$5~c/s) observed in this study are: i) spurious electron emission from the gate grid; ii) delayed electron emission from the liquid surface~\cite{Sorensen17}; iii) interactions occurring in low-PDE regions of the chamber such that the S1 pulse may be missed. Emission rates as low as 10~c/s ($\sim$10$^{-18}$~A) can be detected for the upper wire surface, considering the 50\% detection efficiency. This is a factor of 10$^{10}$ lower than the current resolution of the cathode supply, which never registered any increase, and at least 100$\times$ lower than the regime explored by previous studies based on current measurements rather than electron counting~\cite{Noer82}. In addition, localised bursts spanning consecutive waveforms can be detected with very high significance.

The field on the wire is always minimised until the first ramping test to avoid any conditioning prior to data taking. The first test is conducted in the configuration described in Section~\ref{subsec:Electrostatics}, which allows higher detection efficiency. The test concludes when the cathode supply trips due to the feedthrough limitation mentioned above. Calibration of the SE pulse size distribution is conducted immediately after this first test. Often, a second voltage ramping is conducted using the same field parameters to search for any conditioning effects. Finally, the wire is stressed to higher fields in a third test by grounding the gate, despite this decreasing the detection efficiency.

\subsection{Calibration} \label{subsec:Calibration}

\begin{figure}[h]
\begin{center}
\includegraphics[width=8.6cm]{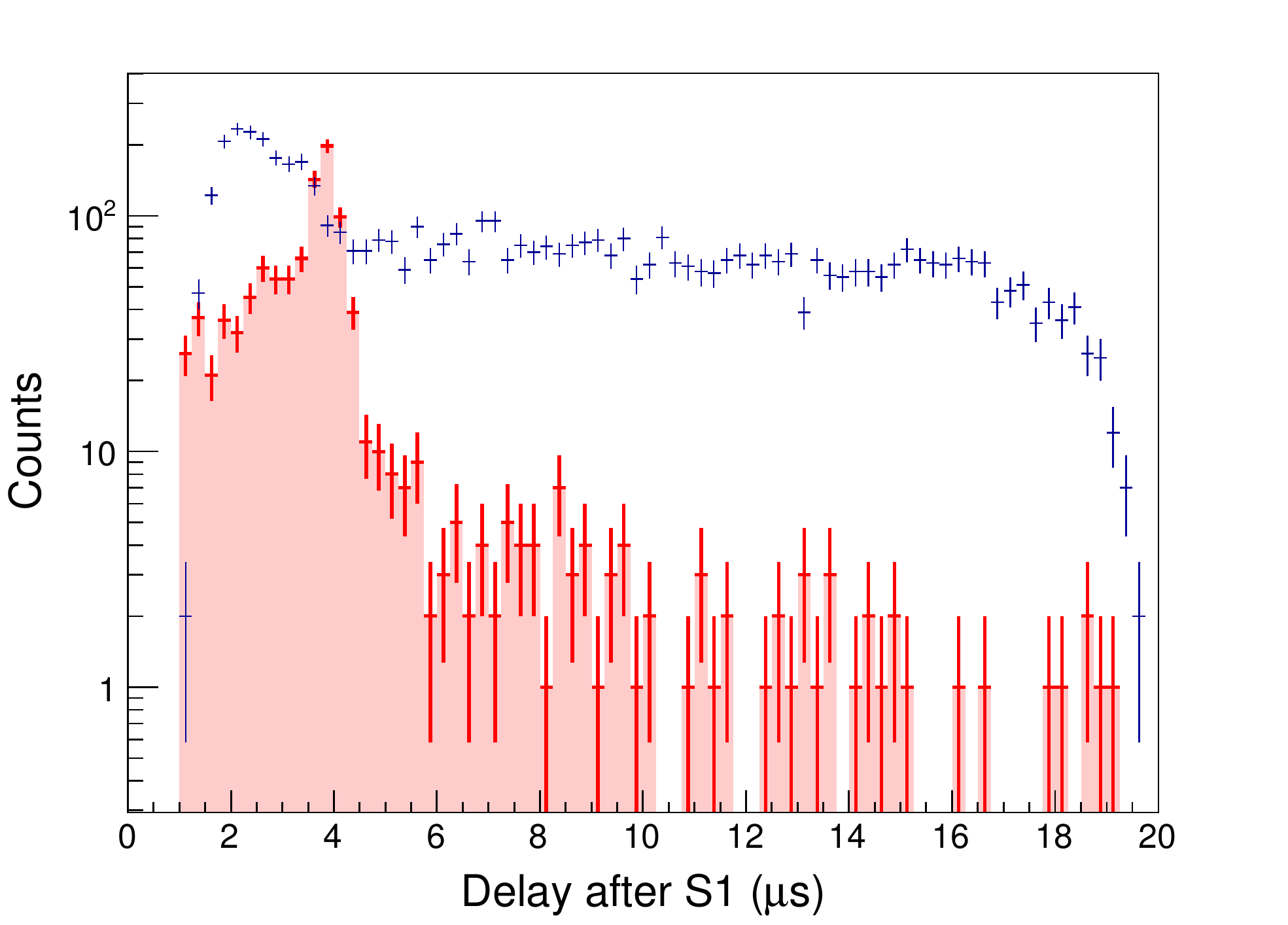}
\end{center}
\caption{Delay time distributions from the S1 response in a calibration dataset: to SE-like pulses in red, and to S2 pulses in blue. Note the peak near 4~$\mu$s due to the gate grid QE, the small population of delayed SE signals mostly due to S1-induced photoionisation, and the small dependence of the S2 rate on drift time confirming high LXe purity.
}
\label{fig:calibration}
\end{figure}

A calibration dataset is acquired after the first ramping test at an intermediate (constant) cathode voltage to establish the SE response in each run. Calibration events are triggered on the PMT signal with a threshold slightly above 1~phe, which allows S1 pulses to trigger the acquisition. SE pulses can be generated by S1-induced photoionisation of impurity species in the liquid, and by photoelectric emission from the gate grid~\cite{Santos11}. The QE for stainless steel at the xenon scintillation wavelength (175~nm) is $\sim\!2\times\!10^{-4}$~\cite{Feuerbacher72}, and so the probability for double phe emission is small. These `gate QE' SE pulses are useful as they occur at a fixed time delay from the S1 optical stimulus---normally $\approx$4~$\mu$s, depending on the particular LXe level in the run---as exemplified in Fig.~\ref{fig:calibration}.

\begin{figure}[thb]
\begin{center}
\includegraphics[width=8.6cm]{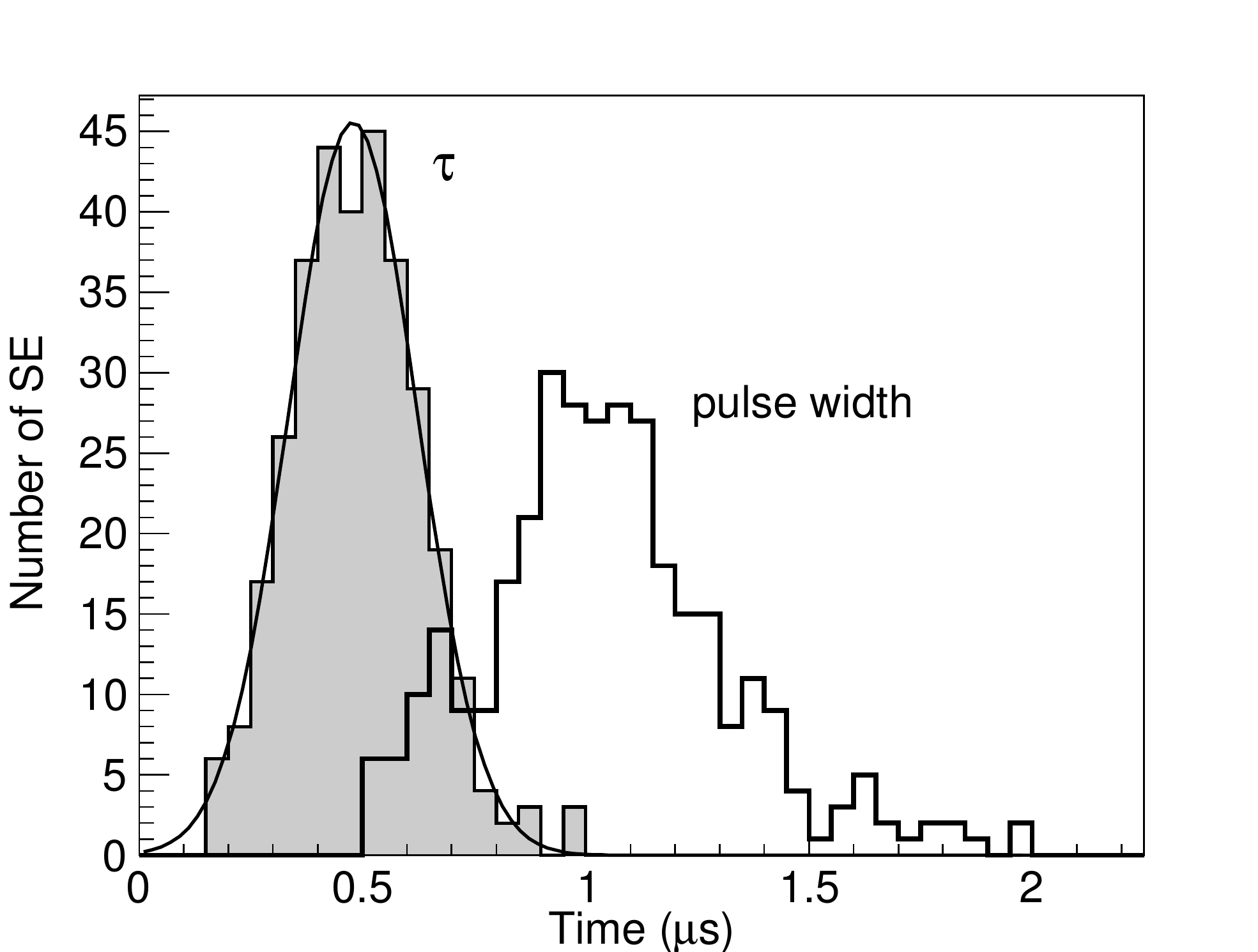}
\end{center}
\caption{SE pulse timing parameters for the gate QE population in a calibration dataset are shown: the parameter $\tau$ (fitted with a Gaussian) and the width at threshold crossing.
}
\label{fig:timecuts}
\end{figure}

A few hundred SE pulses from the gate QE population are used to define the pulse selection criteria for each dataset. These are based mostly on timing cuts since pulse shape features are determined by the gas-phase parameters (density, gap length and field) which are fixed for each run. Example timing distributions are shown in Fig.~\ref{fig:timecuts}. A $\pm2\sigma$ cut is placed on the Gaussian-fitted distribution of the mean arrival time, $\tau$, of the PMT charge measured from the start of the pulse; an additional cut is placed on the total pulse width. The means of these width parameters are consistent with those expected for the drift speed in xenon vapour. No explicit upper cut is applied to the pulse area in order to include multiple-electron emission and considering the position dependence of the S2 PDE discussed in Section~\ref{subsec:Optics}. However, loose cuts are applied to the pulse amplitude to exclude an insidious pick-up observed in our waveforms. The gate QE population is also measured from any S1 signals present in the search data in order to monitor the chamber performance during the ramping tests. These cuts are considered in the calculation of the pulse selection efficiency discussed earlier.

Attention is paid to long drift-time SE and S2 populations as these are sensitive to the electron lifetime in the liquid. The S2 population shown in Fig.~\ref{fig:calibration} is enhanced at short times due to the lower threshold in the strong field above the gate; part of the increased SE rate is instead attributed to the misclassification of S1-induced artefacts such as PMT afterpulsing and ringing. The PMT gain is calibrated from the same data by selecting dark counts preceding S1 pulses.

%%%%%%%%%%%%%%%%%%%%%%%%%%%%%%%%%%%%%%%%%%%%%%%%%%%%%%%%%%%%%%%%%%
\section{Results} \label{sec:results}

Over a 3-year period we tested 18~wire samples with diameters in the range 40--500~$\mu$m, most with well established provenance---with several from the spools used to manufacture the electrode grids for the ZEPLIN-III~\cite{Akimov07} and LUX dark matter experiments~\cite{Akerib13} and the Xed detector~\cite{Shutt07}. The most conclusive LXe runs are summarised in Table~\ref{tab:summarytable}, which lists key wire parameters and the results discussed later in this section. Note that these results pertain to the first (of three) ramping tests unless indicated.

The first stage of the study explored different wire diameters and materials. The 200~$\mu$m LUX cathode wire was studied in greater detail, once it was established as the most prolific for spurious emission and, in the final stage, surface treatments were tested on this wire. We detected event topologies localised in time (and therefore in surface electric field) in all untreated samples that we attribute to electron emission. No macroscopic breakdown was observed from discharging or arcing from any wire, with the tests concluding with a cathode supply trip due to the well-understood feedthrough limitation mentioned previously.

\begin{table*}
\caption{Summary of wire parameters and results for the {\em first} voltage ramping test on each wire (the terminology is explained more fully in the text). $E_{max}$ is the maximum field on the wire surface applied in each test, calculated for a perfect cylinder; `H/L' denotes high/low sensitivity in test, i.e.~gate at $-$2/0~kV, respectively. `FE' represents the average rate of faint emission. EE ($E_{min}$) is the number (and lowest field in kV/cm) of extended emitters. `IE M/C' indicates the number of impulsive emitters with multiple SE pulses within a waveform (M) or in consecutive waveforms (C).}
\label{tab:summarytable}
\renewcommand{\arraystretch}{1.17}
%\centering
\begin{center}
\begin{threeparttable}

\begin{tabular}{c|lcc|c||ccc}
%Run & Wire/Treatment & \diameter ($\mu$m) & Material & Max $E$ H/L \tnote{a} (kV/cm) & Avg (Hz) & FE \tnote{b} (min $E$) & M/C \tnote{c}\\
\hline
\hline
Run & Wire/Treatment & \diameter ($\mu$m) & Material & $E_{max}$ H/L (kV/cm) & FE (c/s) & EE ($E_{min}$) & IE M/C \\
\hline
10\tnote{a} & Xed                & 40 & BeCu  & 310/405\tnote{b}  & $(<\!210)$       & none   & n/a \\
\hline
12\tnote{a} & LUX gate                & 101.6 & SS304  & 123/152   & $(<\!113)$       & 1 (107)\tnote{c}   & n/a \\  
14 &                         &  &  & 122/160   & $52\pm4$       & 3 (20)   & 31/2 \\  
\hline
15 & LUX cathode             & 205.6 & SS302  & 70/84     & $45\pm3$       & 2 (10)   & 26/4 \\ 
16 &                         &     &        & 69/91     & $19\pm2$       & 2 (55)   & 6/2 \\ 
26\tnote{d} &                         &     &        & 86/102    & $23.5\pm1.8$   & none     & 3/0 \\
19 & \quad electropolished    &     &        & 66/92     & $15.6\pm1.7$   & none     & 3/1 \\ 
25 & \quad acid-cleaned       &     &        & 65/94     & $2.8\pm0.7$    & none     & 0/0 \\ 
27\tnote{d} & \quad\quad aged, 1$^{\mathrm st}$ run    &     &        & 70/104    & $3.8\pm0.8$    & none     & 0/0 \\ 
29\tnote{d} & \quad\quad aged, 3$^{\mathrm rd}$ run\tnote{e}      &  &  & 67/108    & $5.3\pm0.9$    & none     & 0/1 \\ 
\hline
17 & ZEPLIN-III cathode           & 99.1 & SS316L & 119/161   & $27\pm2$       & none     & 2/0 \\ 
20 & \quad electropolished        &     &        & 119/161   & $5.8\pm1.0$    & none     & 4/0 \\ 
\hline
22 & Gold-plated tungsten    & 125 & W(Au)  & 96/134    & $11.7\pm1.4$   & none\tnote{f}     & 2/0 \\
\hline
\hline
\end{tabular}

\begin{tablenotes}
	\footnotesize
	\item[a]{Runs with non-optimal DAQ settings (shorter waveforms, higher SE backgrounds) indicated in brackets---see text.}
    \item[b]{Sensitivity only to light emission.} 
    \item[c]{Largest emitter observed: 3,000~c/s bin-average.}
    \item[d]{Runs at 2.45~bar xenon pressure.}
    \item[e]{Runs 27--29 same wire bagged in air for 8~months after acid-cleaning; tested 3 times and stored in vacuum between runs.}
    \item[f]{An extended emitter was observed during the third ramping test (low sensitivity settings) at 99~kV/cm.}
\end{tablenotes}

\end{threeparttable}
\end{center}
\end{table*}

In this section we begin by defining our emission terminology. We then describe preliminary measurements done prior to optimising the system for emission rate sensitivity (Runs~$\le$12)---as some interesting conclusions can be drawn from those earlier tests, and they include the largest emitter we recorded in this study as well as the highest electric field reached. At this point we summarise the evidence to support that this phenomenology is indeed consistent with spurious electron (and photon) emission from our samples. We then describe in more detail some tests conducted with the improved sensitivity, before presenting results for the various wire types and surface treatments.

\begin{figure*}[tbh]
\begin{center}
\includegraphics[width=18cm]{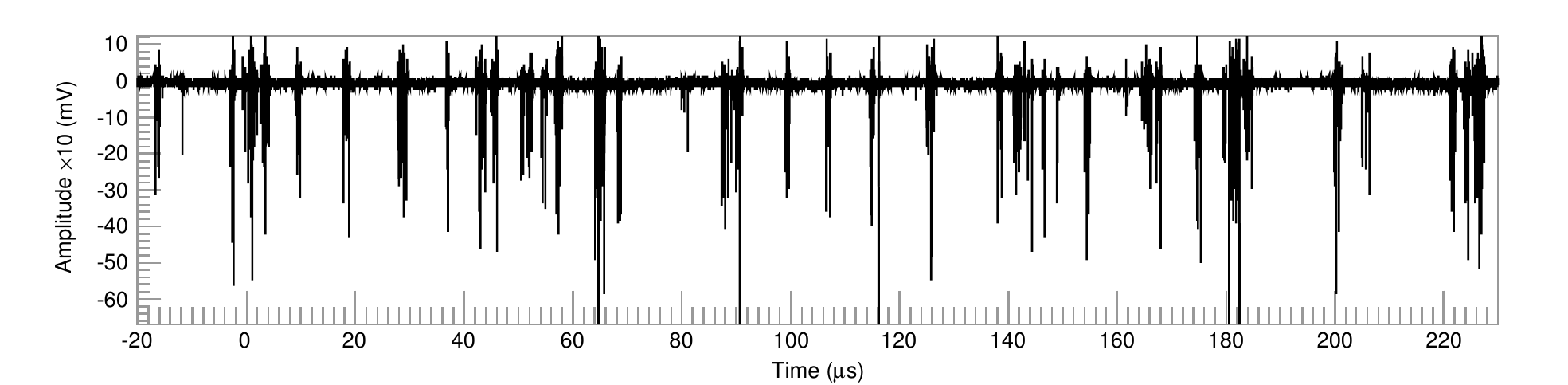}
\end{center}
\caption{Multi-candidate 250-$\mu$s waveform containing 35 SE pulses (instantaneous rate $>$100,000~c/s) registered at 58.7~kV/cm in the first voltage ramping test of Run~15, cf.~Fig.~\ref{fig:Run15_rates}. The previous two waveforms contained one and three candidates for a 0.275~s total elapsed time. This is the most intense impulsive emitter we recorded in our study.
}
\label{fig:MultiCandidateEvent}
\end{figure*}

\begin{figure}[tbh]
\begin{center}
\includegraphics[width=8.6cm]{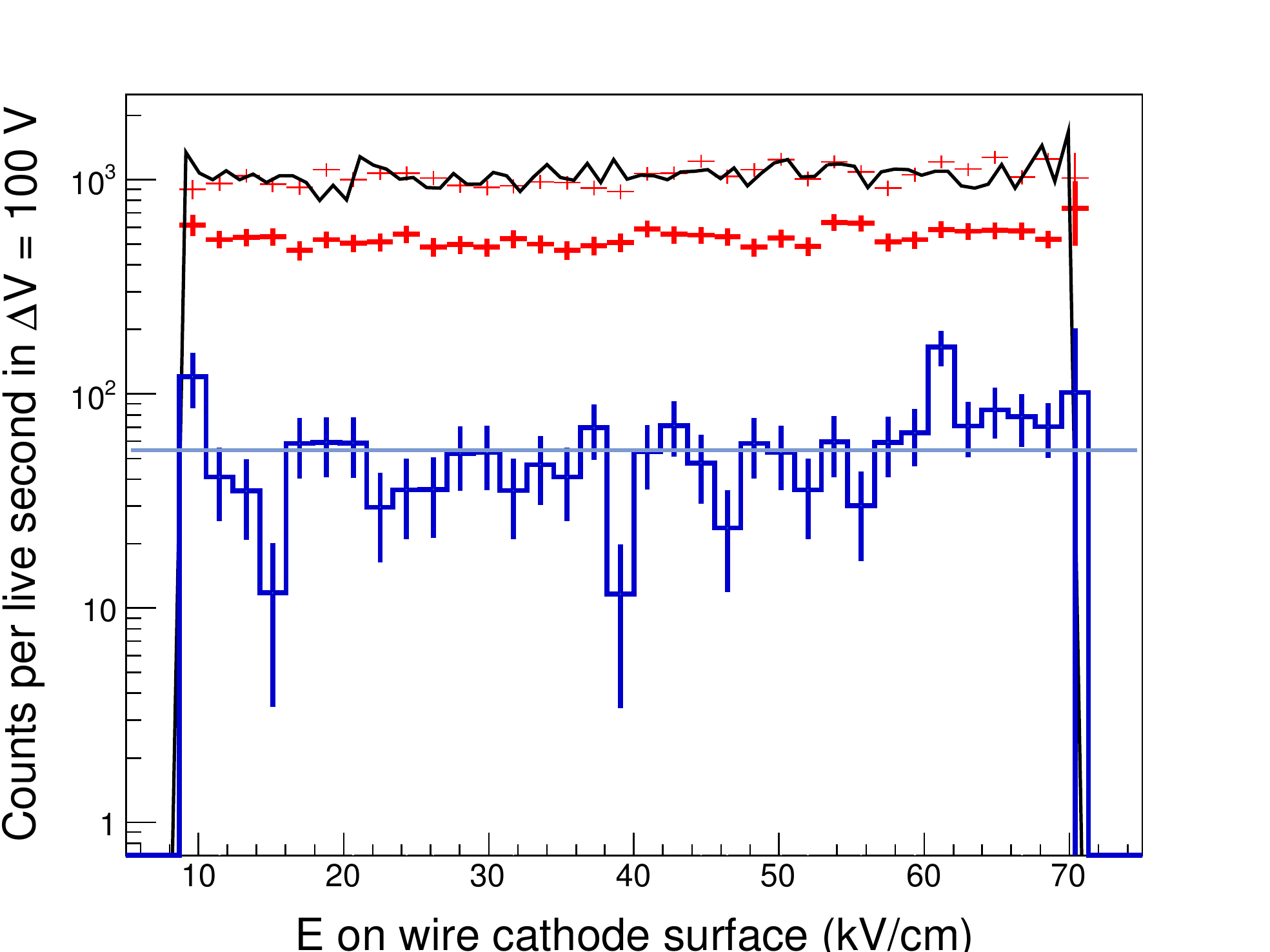}
\end{center}
\caption{First voltage ramping results for a 200-$\mu$m LUX cathode wire (Run~15): emission rate as a function of field on the wire surface (blue), with the average indicated by the horizontal line; single phe pulses (thin red markers); S1-type signals, i.e.~fast pulses with area $\ge$3~phe (thick red markers); all pulse types (black, divided by 10).}
\label{fig:Run15_rates}
\end{figure}

We define the terminology used in this study for the different manifestations of electron emission with reference to Figs.~\ref{fig:MultiCandidateEvent} and~\ref{fig:Run15_rates}, relating to a typical voltage ramping test. We classify emission in three categories, based both on emission rates and the range of fields over which it occurs. From the most to the least pronounced:
\begin{enumerate}
\item {\em Extended emitters} (EE in Table~\ref{tab:summarytable}) are significant enhancements of emission rate which persist over a measurable range of fields ($\gtrsim$1~kV/cm); note a very clear extended emitter around 60~kV/cm in Fig.~\ref{fig:Run15_rates}, and a weaker one at 10~kV/cm.
\item {\em Impulsive emitters} (IE) are much faster but even more intense ones which appear and disappear in a short period of time (such that the field variation is negligible); Fig.~\ref{fig:MultiCandidateEvent} displays an extreme example.
\item {\em Faint emission} (FE) refers to the low, but relatively steady and field-independent emission rate; this is indicated by a horizontal line at $\simeq\!50$~c/s in Fig.~\ref{fig:Run15_rates}.
\end{enumerate}

In general, for all untreated samples we observed faint emission at all fields, and this rate is not compatible with the low background of the measurement ($\lesssim5$~c/s, as shown later in this section). The faint emission rate indicated in Table~\ref{tab:summarytable} is estimated from a likelihood fit of a constant rate over the whole voltage ramping test, excluding only the very obvious emitters (i.e.~excursions larger than $3\sigma$ or consecutive bins with excursions larger than $2\sigma$). We found also many examples of extended emitters, with peak intensities varying by orders of magnitude; these are defined as $>\!2\sigma$ excursions above the faint emission rate over a single 100~V bin, corresponding to a narrow range of field. However, impulsive emitters were far more common; these were detected from the presence of multiple candidate pulses within a single waveform or across a few adjacent waveforms, and these two types are indicated separately in the table.

\subsection{Electron emission in early tests}

\begin{figure}[tbh]
\begin{center}
\includegraphics[width=8.6 cm]{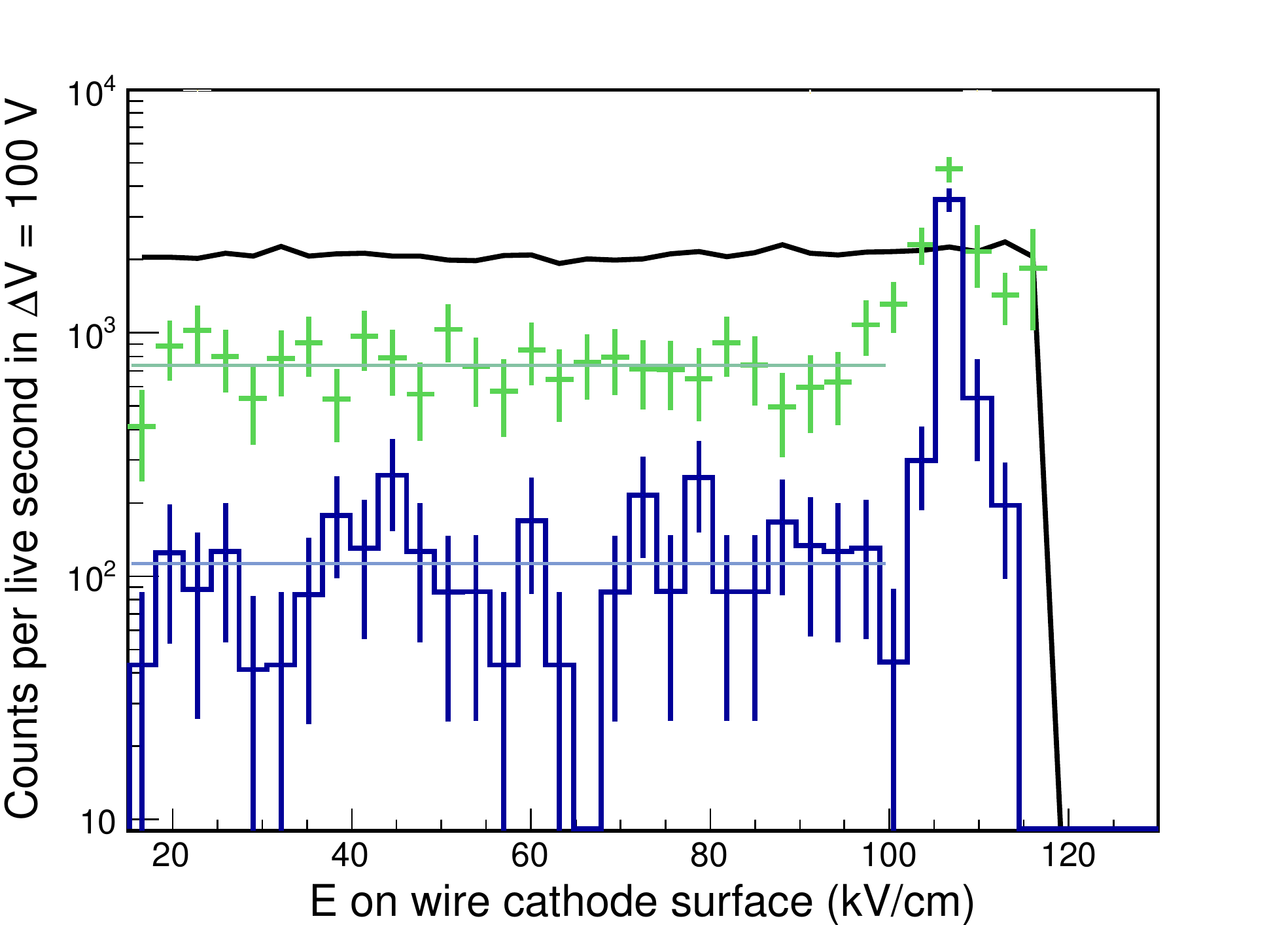}
\end{center}
\caption{Run~12 test of a 100~$\mu$m LUX gate wire, which revealed the strongest emitter in our study. The plot shows the candidate pulse rate as a function of field (blue); post-S1 SE pulses, typically from gate QE and photoionisation (green markers)---this population increases during the emission episode due to SE candidate pulse leakage into that time window; no effect is noticeable on the rate of all pulses (black, divided by 10), which is instead very sensitive to photon emission from the feedthroughs.
}
\label{fig:ignition}
\end{figure}

Before acquisition settings were fully optimised for live time and SE background, several wire measurements were conducted and useful results derived (Runs~$\le$12). For example, we were able to observe an extended emitter so intense that it led to a detectable current in the PMT power supply. This emitter, comparable in rate to those studied in LUX from its gate grid (see Ch.~6 in Ref.~\cite{AdamThesis}), was recorded when testing a sample of the same wire. As shown in Fig.~\ref{fig:ignition}, the ignition persisted from 104 to 110~kV/cm, averaging $\sim$3,000~c/s. On a finer time resolution the instantaneous rate is observed to build up and then decline. Most waveforms collected during this emitter contain only one SE pulse, yet multiple-candidate waveforms were also found which represent a peak rate of $\sim$500,000~c/s. The same sample did not emit again when tested up to 152~kV/cm with the gate at ground (such an emitter would have been easily detected despite the lower sensitivity) and it held 255~kV/cm with no photon emission. Previously, two samples of 100~$\mu$m stainless steel wire used in ZEPLIN-III had achieved fields in excess of 160~kV/cm with no signs of major emission (not listed in the table). Still with these preliminary settings, the 40~$\mu$m BeCu wire used in the Xed detector~\cite{Shutt07} was shown to sustain fields as high as 310~kV/cm without detection of any major emission, and it successfully held 405~kV/cm with sensitivity only to light. 

\begin{figure}[tbh]
\begin{center}
\includegraphics[width=8.6cm]{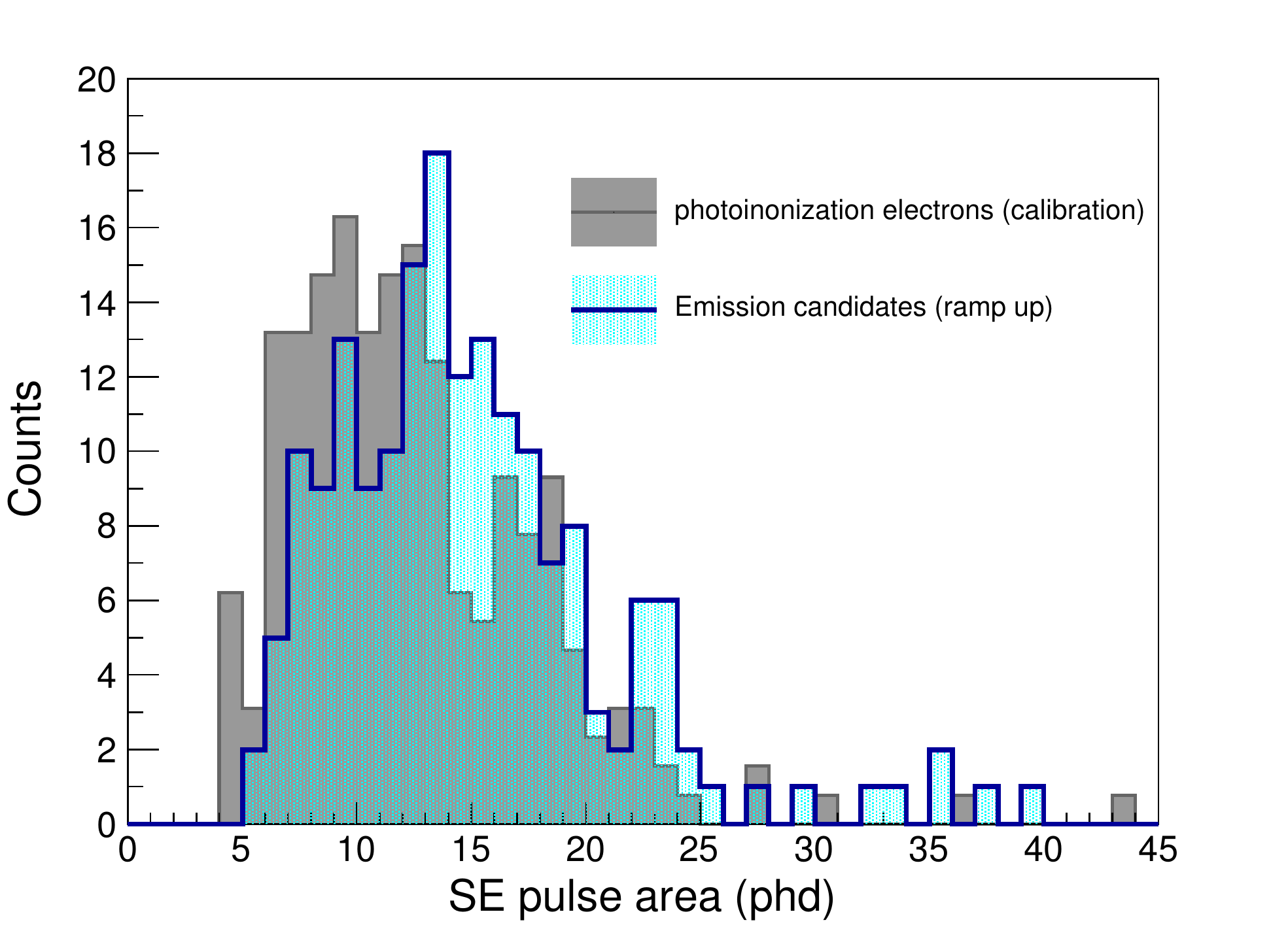}
\end{center}
\caption{Pulse size distribution for emission candidates (180 pulses from the ramping test in Run~16) compared with that obtained from gate-grid QE pulses (232 pulses selected from the calibration dataset for the same run, as described in the text). The small shift of $\sim$10\% observed here occurs in all runs.
}
\label{fig:SEsizes}
\end{figure}

There are several reasons to attribute the phenomena discussed so far to electron emission from our samples. Firstly, we established that indeed this emission consists of single electron pulses, as illustrated in Fig.~\ref{fig:SEsizes}, where we overlay the size distribution for a population of SE pulses from the gate QE calibration with emission candidate pulses. They are consistent, but not identical---and this is anticipated due to the different PDEs expected for both populations (originating on a disc as opposed to a line). For the same reason, if these signals came from elsewhere in the chamber, a reduction of pulse size would have been expected instead. We can be confident that this emission does not arise in the gate grid. The frequency of intense and extended emitters increases generally with the cathode wire field---while the surface field on the gate wires ($\sim$15~kV/cm) actually decreases somewhat for higher cathode voltages. In addition, the production of light in coincidence with charge should have been much more easily detected from the gate than from the cathode wire. We discuss these topics below in more detail. Finally, the clear reduction in emission following treatment of the cathode wires confirms that the gate is not implicated.

\begin{figure}[tbh]
\begin{center}
\includegraphics[width=8.1cm]{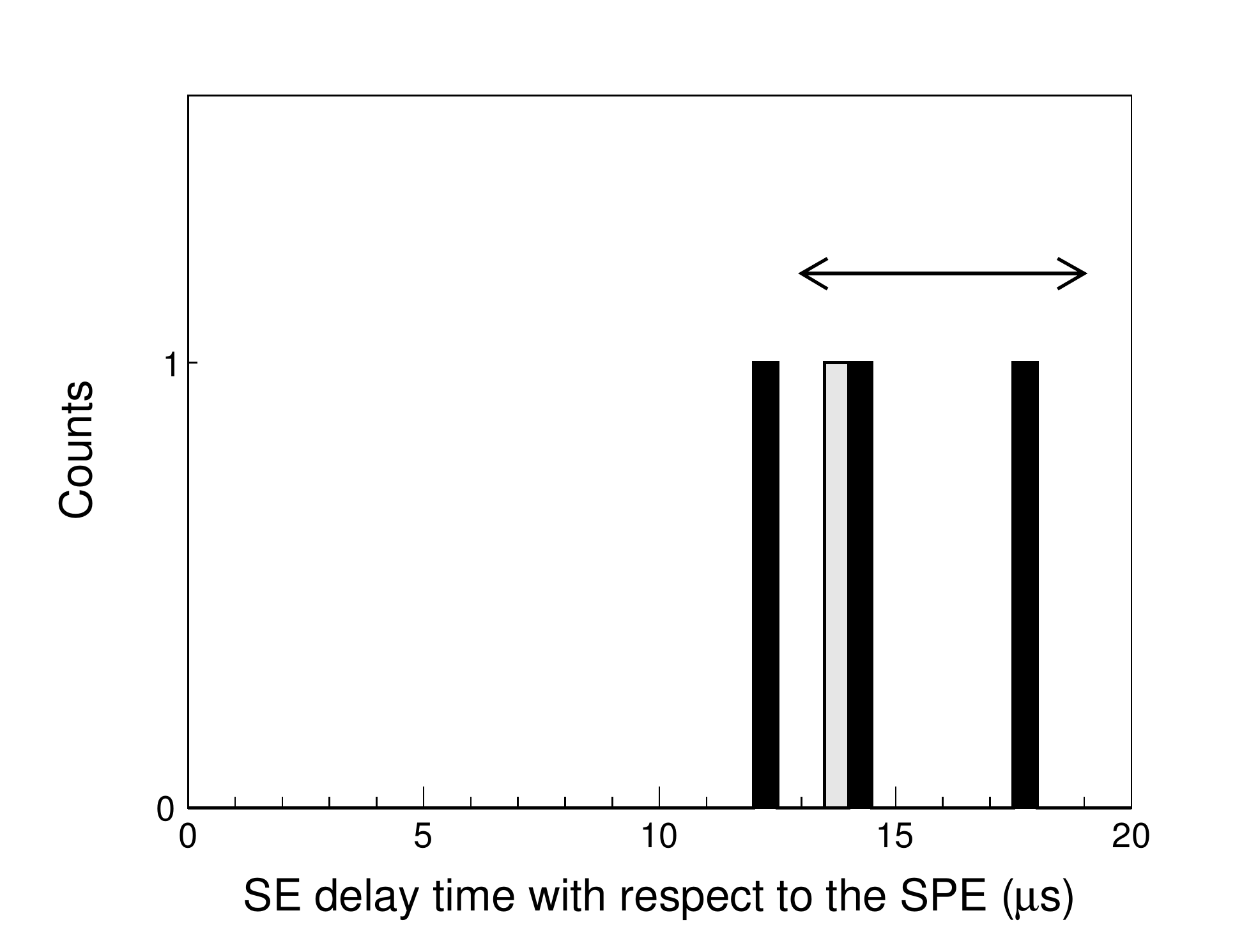}
\end{center}
\caption{Coincident photon and electron emission observed during the large extended emitter in Run~12 (black), and over the remainder of the test (grey). The arrow indicates the range of drift times for electrons released from the wire at this field, which has a mean value near 15~$\mu$s.
}
\label{fig:lightEmission}
\end{figure}

In general, we cannot determine that each electron originates at the wire directly from its drift time---as one would for a typical energy deposition in a LXe-TPC, i.e.~by the time separation between the light (S1) and the ionisation (S2) responses---unless photon production also occurs simultaneously. In fact, such coincident emission of electrons and photons was detected in LUX~\cite{AdamThesis}. In our case: i) the chamber has very low PDE for photon emission in the liquid, as shown in Fig.~\ref{fig:GEANT4}; ii) the drift paths and fields along them depend significantly on the emission angle from the wire, as highlighted in Fig.~\ref{fig:COMSOL}; and iii) the field is not constant during the test and therefore neither is the time delay between any optical signal and the SE detection.\footnote{All three factors would make identification of any emission from the gate grid more favourable: the photon PDE is higher and the drift times are constant in this instance. Our results are not consistent with significant emission from the gate wires.} However, we were able to find evidence for such photon emission from our largest emitter described previously (Run~12), for which the difficulties just exposed are partially avoided. Figure~\ref{fig:lightEmission} shows single phe detections found in a time window consistent with the drift time from the sample at that particular field. This contrasts with the delay time distribution between S1 pulses and photoionisation electrons which is known from calibrations. Four single phe pulses were identified preceding a total of 552 emission candidates; assuming the PDE estimated in Section~\ref{subsec:Optics}, we may conclude that the number of photons emitted in this instance is comparable to the number of electrons---also found to be the case in Ref.~\cite{AdamThesis}.

\subsection{Results from high sensitivity tests}

A typical first-ramping result obtained now with the improved sensitivity afforded by the longer waveforms was that shown in Fig.~\ref{fig:Run15_rates} for the 200-$\mu$m LUX cathode wire (Run~15). Two extended emitters are visible, one at $\sim$10~kV/cm and the other at $\sim$60~kV/cm, above a rather constant faint emission rate of 45~c/s. The analysis confirms that only the candidate rate increases, while that for the other pulse types remains constant. This type of emitter persists for up to a few minutes---corresponding to a narrow range of fields---before declining. During their active period, a detailed electric field dependence could not be established (e.g.~a Fowler-Nordheim law characteristic of field emission~\cite{FN28}) due to the externally-triggered acquisition and the relatively low emission rate.

\begin{figure}[tbh]
\begin{center}
\includegraphics[width=8.6cm]{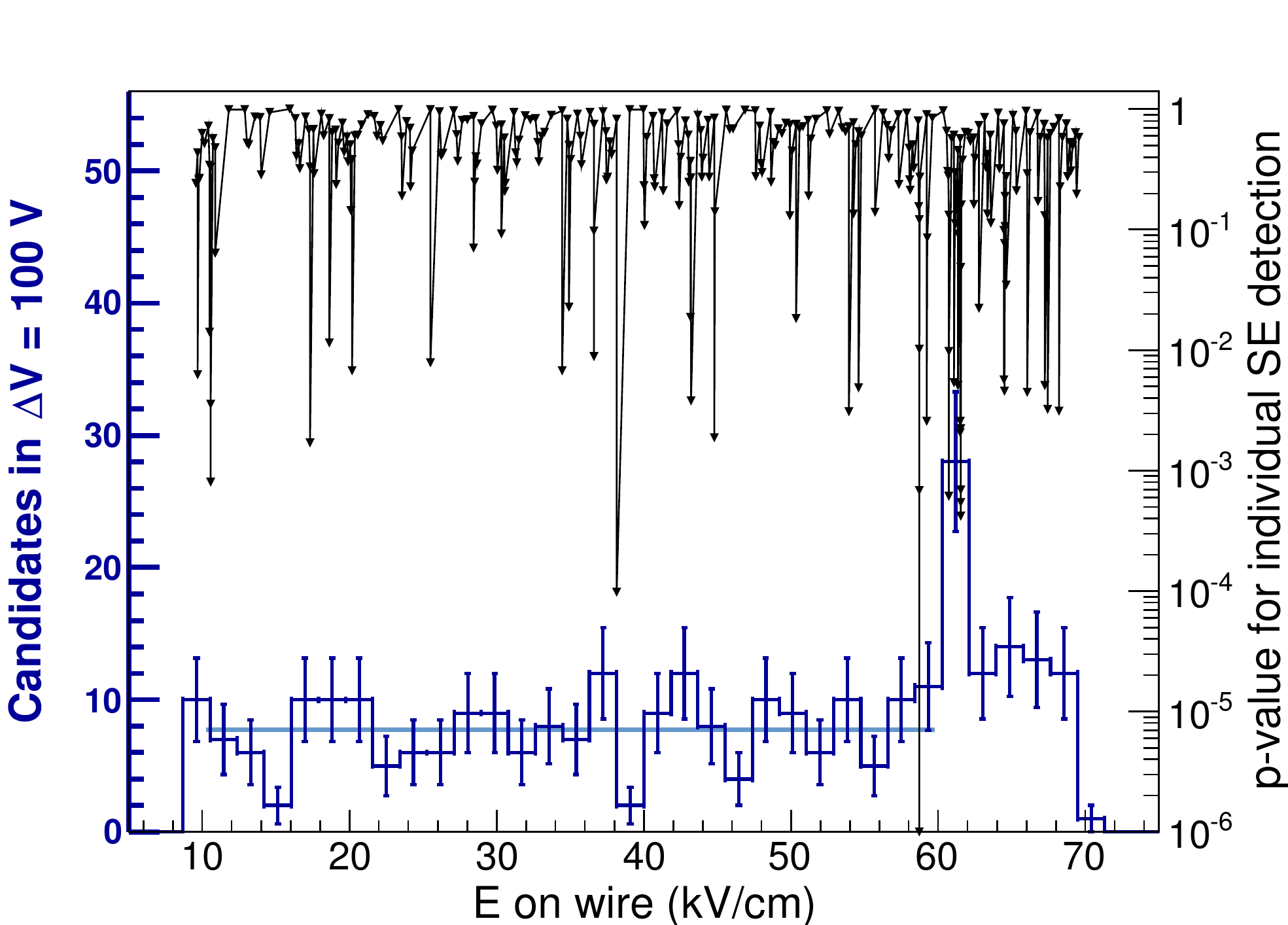}
\end{center}
\caption{Counts per 100-V bin for the Run~15 data shown in Fig.~\ref{fig:Run15_rates} (histogram); the upper axis shows the elapsed-time $p$-value for each individual SE detection.}
\label{fig:Run15_pvalues}
\end{figure}

On closer inspection, the waveforms which contribute to these extended emitters show substructure, sometimes with instantaneous rates peaking at $\sim$100,000~c/s before subsiding quickly, all during a fraction of a second---similar to the impulsive emitters defined above. These enhancements are not, for the most part, stochastic excursions expected above a constant emission rate. This point is illustrated in Fig.~\ref{fig:Run15_pvalues} for the same data, where the total counts per 100~V bin are plotted with Poisson errors. Above it, we indicate a $p$-value for the time elapsed since the previous SE detection, i.e.~the probability for this time to be shorter than that observed given a Poisson process with a mean rate equal to the faint emission rate. This helps identify individual impulsive emitters. This figure depicts a more complex emission pattern during the test: while it unveils the presence of clear impulsive events within the extended emitters themselves, other detections with significant $p$-values occur in otherwise quiet regions. In one notable instance---the waveform shown in Fig.~\ref{fig:MultiCandidateEvent}---the $p$-value is minuscule; that was the most intense impulsive  emitter we observed.

In Table~\ref{tab:summarytable} the number of impulsive emitters is given both when multiple electrons occur in a single waveform (type M) or in consecutive waveforms (type C). Typical $p$-values for these entries are $\lesssim10^{-4}$ for untreated wires (i.e.~$\sim$1 expected per test).

Still regarding the same test in Fig.~\ref{fig:Run15_rates}, there is an extended emitter in the first bin at 10~kV/cm which appears at first to have low significance; however, the more detailed analysis in Fig.~\ref{fig:Run15_pvalues} confirms several events with low $p$-value (one M- and two C-type emitters). Therefore, there is significant observation of charge emission at a field as low as 10~kV/cm in this wire.

\begin{figure}[thb]
\begin{center}
\includegraphics[width=8.6 cm]{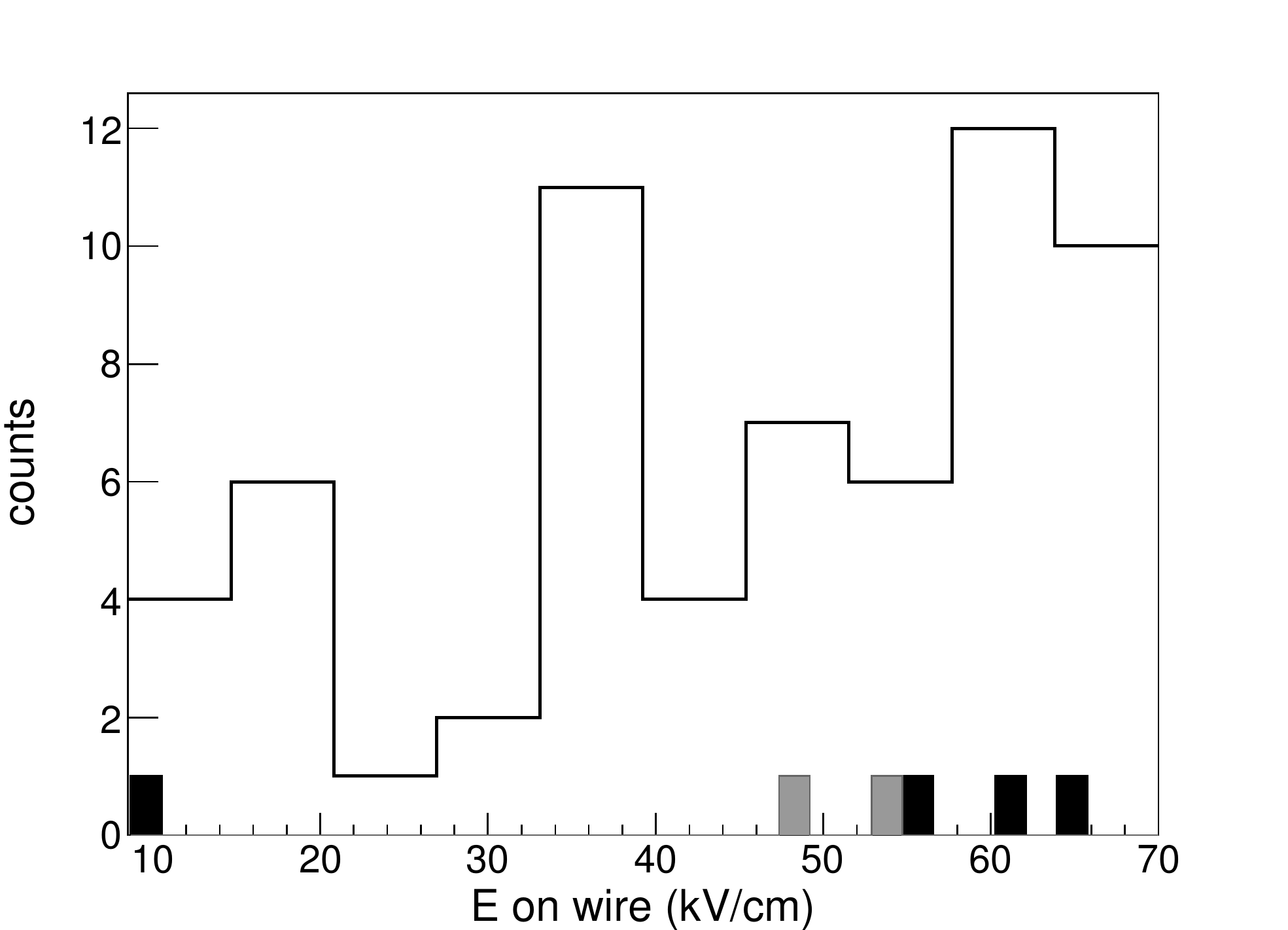}
\end{center}
\caption{Field distribution of impulsive (line histogram) and extended emitters (solid histograms) accumulated for the 200~$\mu$m wire samples during the first (black) and second (grey) ramping tests. The prevalence of impulsive emitters near 35~kV/cm is mainly due to the second ramping test of Run~15. Except for this particular excess, a similar distribution is obtained for the 100~$\mu$m wires.}
\label{fig:emittersDistribution}
\end{figure}

\begin{figure}[thb]
\begin{center}
\includegraphics[width=8.6cm]{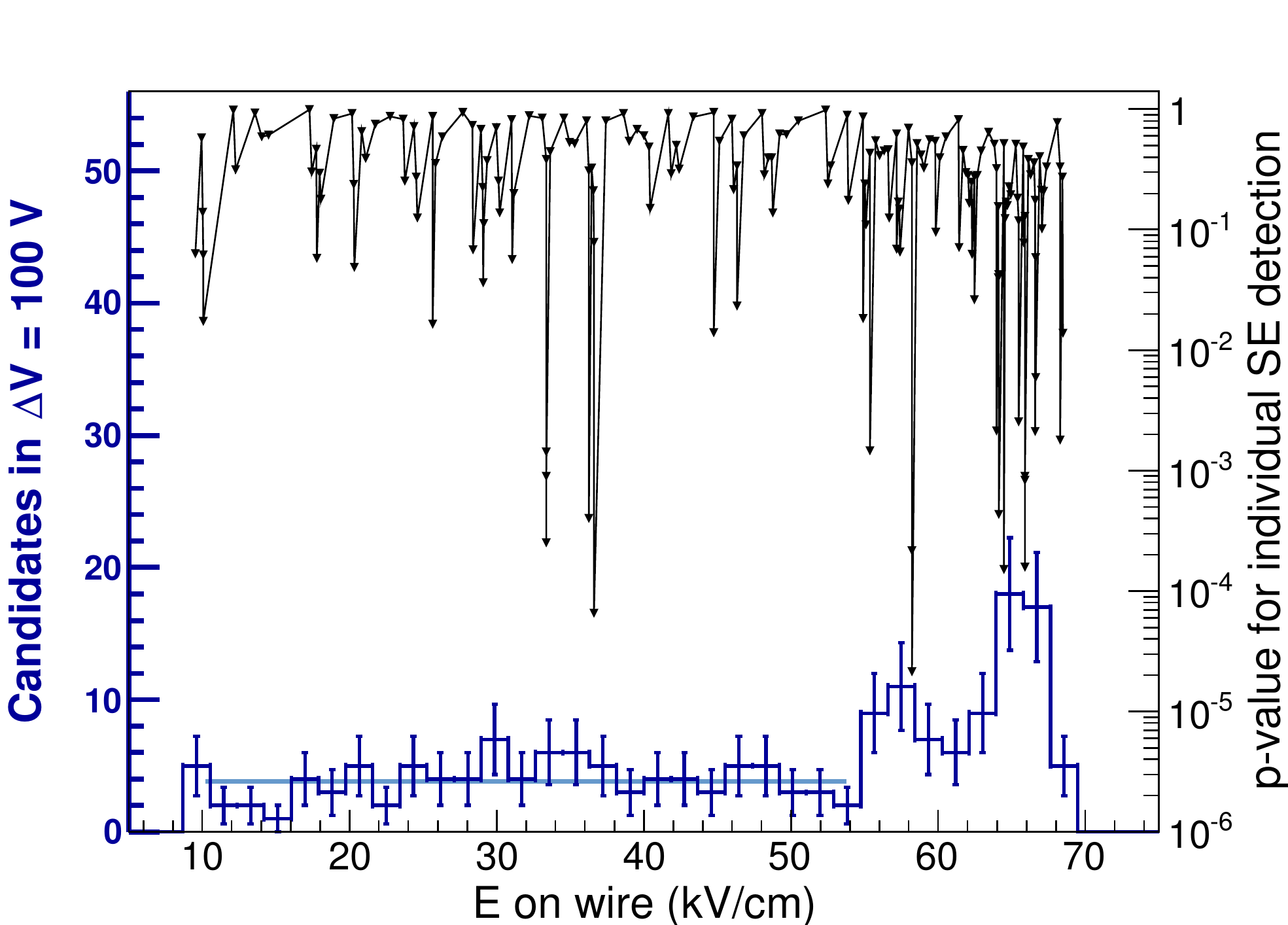}
\includegraphics[width=8.6cm]{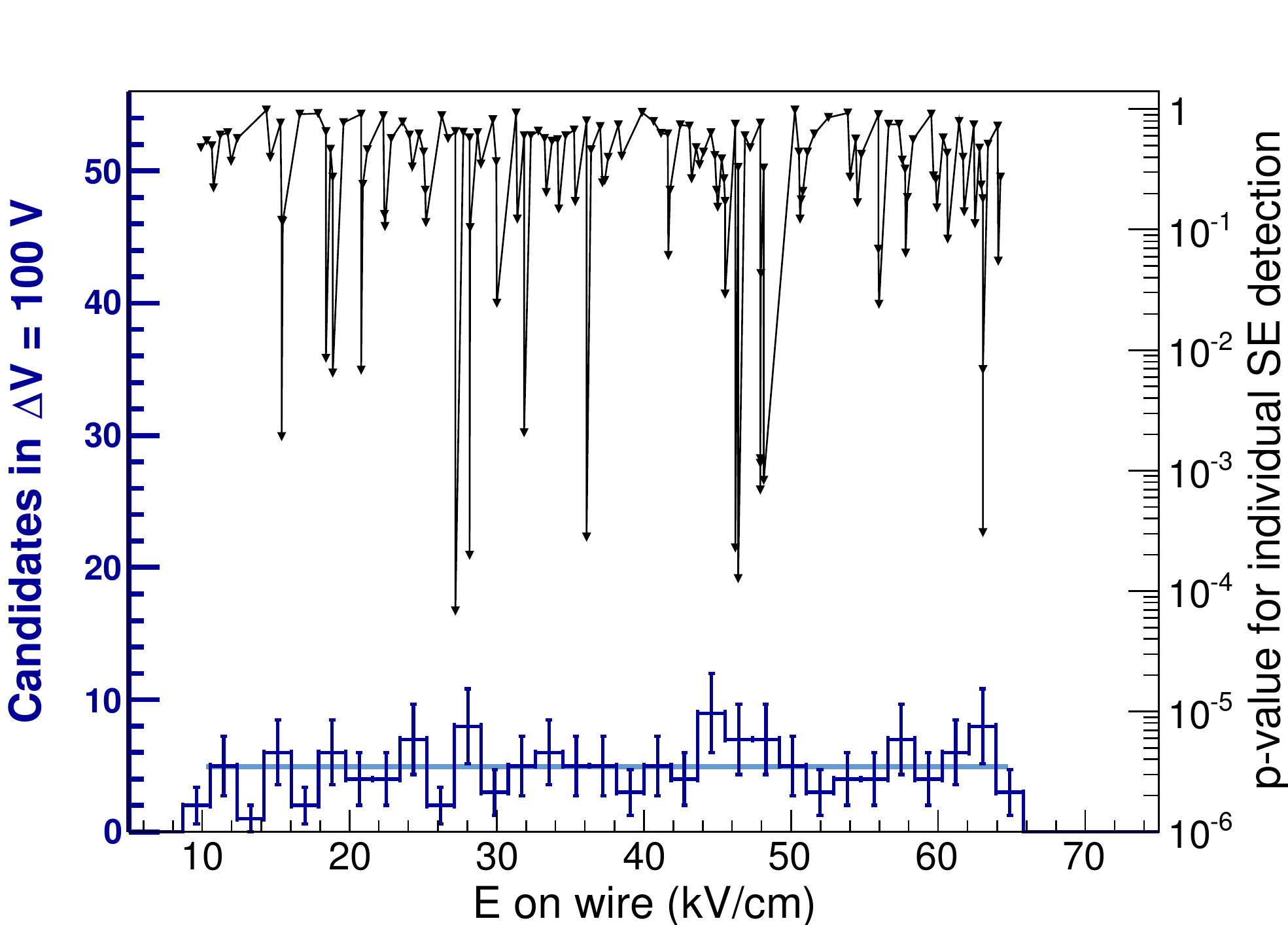}
\end{center}
\caption{Consecutive voltage ramping tests on 200-$\mu$m LUX cathode wire (Run~16); Up -- First ramping; Down -- Second ramping. The third ramping test with the gate at ground (not shown) did not expose any extended emitter.}
\label{fig:Run16}
\end{figure}

This type of behaviour is seen in most tests, with the number of extended emitters varying between 0 and $\sim$4 and several impulsive emitters likely to be found inside each extended emitter or scattered at other fields, although both type of emitters are more common at high fields; this is shown in~Fig.~\ref{fig:emittersDistribution}. Remarkably, the emission patterns are never reproduced in subsequent ramping tests: fewer and smaller emitters are found at significantly different fields, always $\gtrsim50$~kV/cm. A typical example comes from another sample of the same wire (Run~16) in~Fig.~\ref{fig:Run16}.

\subsection{Dependence on stainless steel type}

\begin{figure*}[tbh]
\begin{center}
\includegraphics[width=8.6cm]{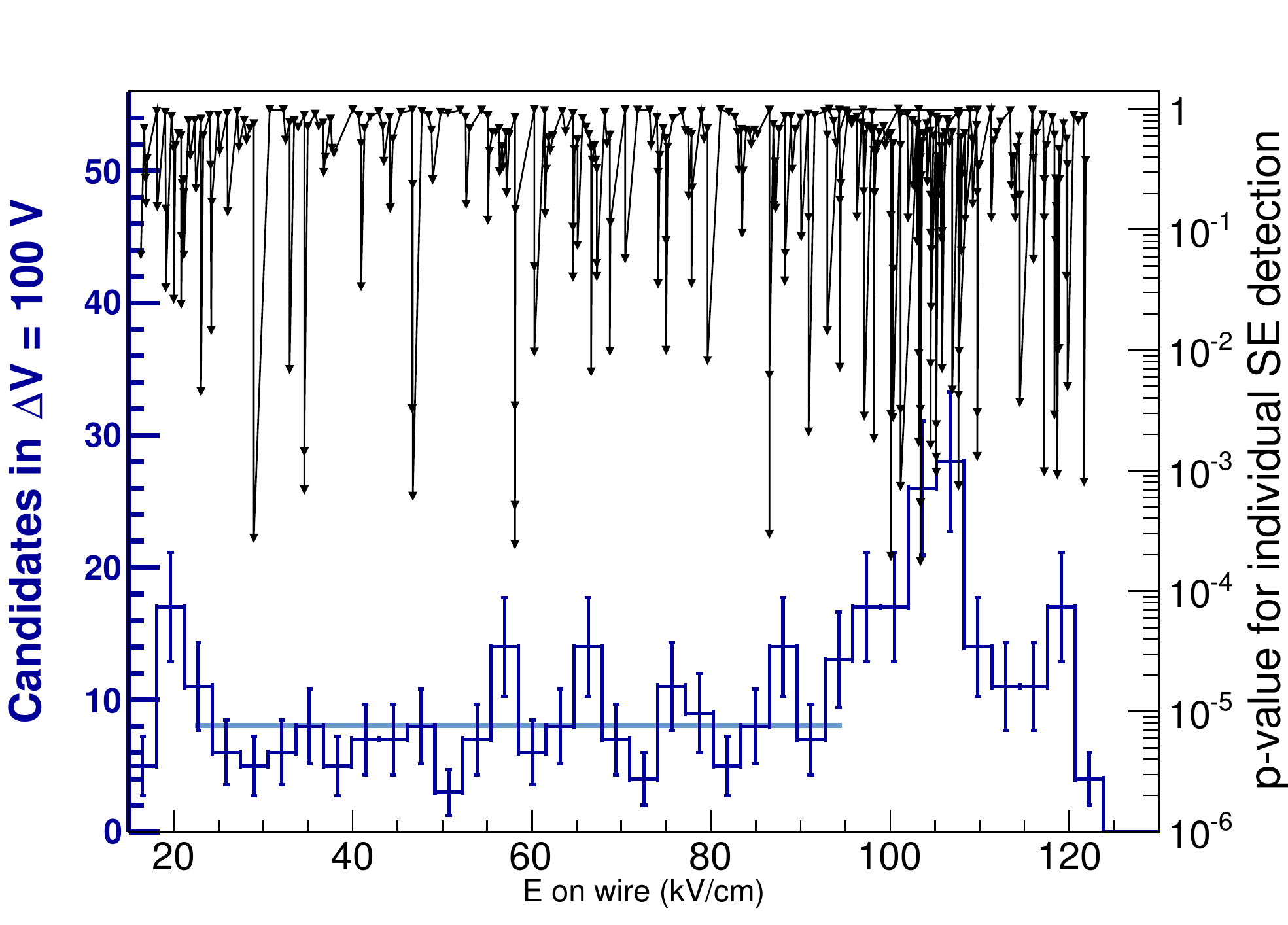}
\includegraphics[width=8.6cm]{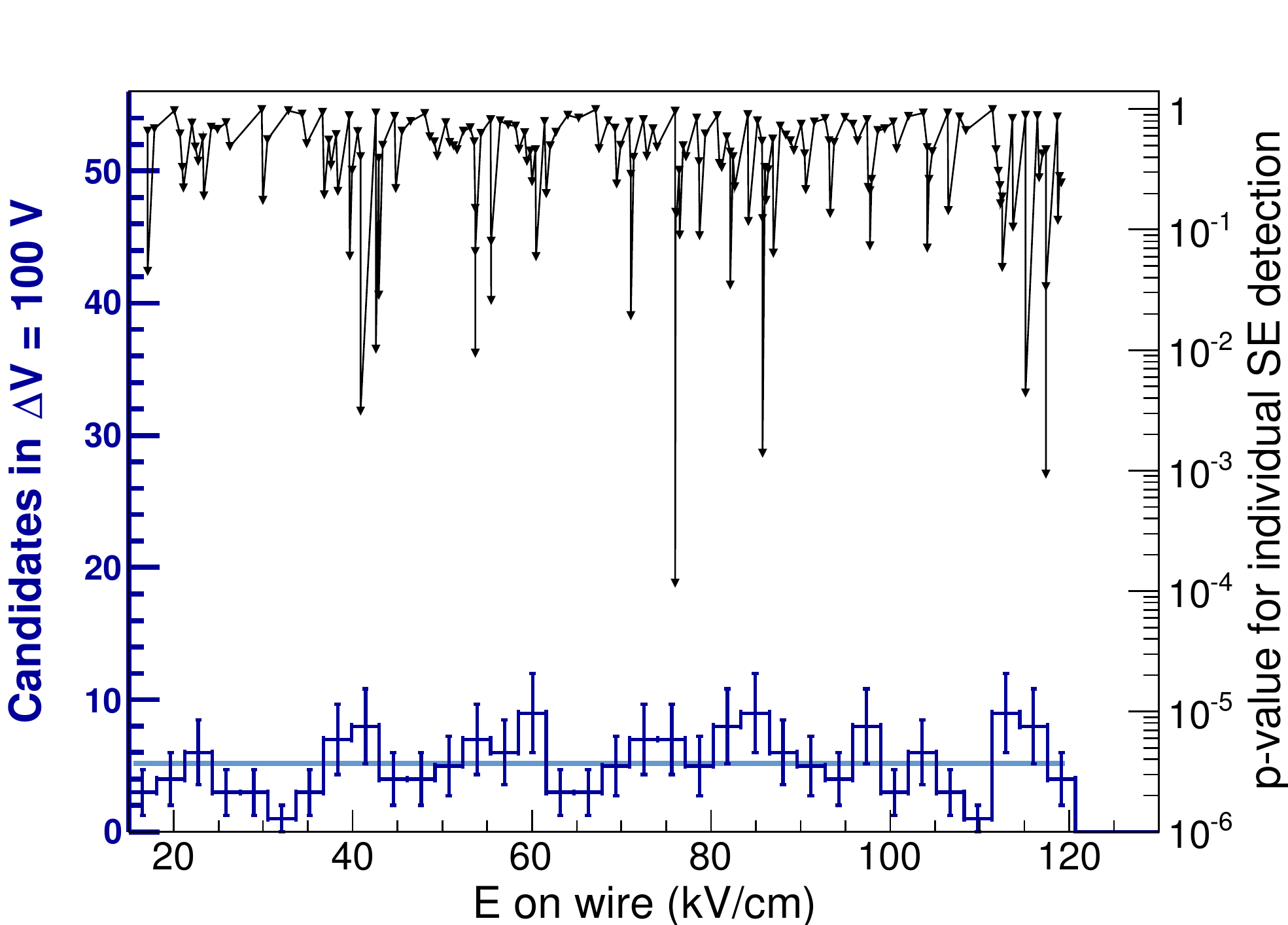}
\end{center}
\caption{(First voltage ramping tests for 100~$\mu$m wire samples. Left -- SS304 LUX gate wire (Run~14); Right -- SS316L ZEPLIN-III wire (Run~17). Compare with the SS302 200~$\mu$m wire samples shown in Figs.~\ref{fig:Run15_pvalues} and~\ref{fig:Run16}, noting the lower scale of applied fields in the latter case.
}
\label{fig:moreSSwires}
\end{figure*}

Besides the 200~$\mu$m LUX cathode wire (SS302), several 100~$\mu$m samples of other stainless steel varieties were tested. Figure~\ref{fig:moreSSwires} (left) shows one of the first voltage ramping tests with optimised sensitivity settings (Run~14) with the LUX gate wire (SS304). The behaviour is similar to that of the SS302 wire, but for a considerably higher surface field. This observation agrees with the experience of the LUX collaboration: a new cathode plane using the 200~$\mu$m wire replaced the initial one made with the 100~$\mu$m wire in an attempt to reduce the surface fields and so improve the HV performance~\cite{LUXr2}, but the improvement was minimal. Indeed, this same 100-$\mu$m wire exhibited the major emitter in Run~12 documented in~Fig.~\ref{fig:ignition}, and similar emission was reported by LUX from its gate grid~\cite{AdamThesis}. Several SS316L wire samples from a spool used in the ZEPLIN-III detector were tested too. Much quieter behaviour was registered with a sample of this SS316L material (known for its anti-corrosion properties), with no observation of extended emitters, and yielding the first impulsive emitter at the relatively high field of 76~kV/cm (Run~17).

\subsection{Effect of surface treatment}

The LUX cathode wire samples provided clear evidence for spurious emission at relatively low fields. Two out of three untreated samples showed multiple extended emitters, and a third exhibited also high faint emission rates and impulsive emission at low fields. Hence, this wire was our reference to test the effectiveness of two types of surface treatment---despite it already being labelled as `ultra-finish' by the manufacturer (California Fine Wire). The treated samples were new, different from those tested previously, but handled and prepared in the same way.

\subsubsection{Electropolishing}

Electropolishing is a common procedure to improve the smoothness of metallic surfaces, and it is particularly effective in reducing protruding defects such as filaments or die marks (the latter were actually visible in SEM pictures of the LUX cathode wire). The electropolishing procedure conducted for our wire samples lasted for 30~s in a Power-Clean~500 solution at $43^\circ$C. The rinse included a brief immersion in 10\% nitric acid at room temperature to remove the electropolishing film, followed by passivation with deionized water.

\begin{figure*}[tbh]
\begin{center}
\includegraphics[width=8.6cm]{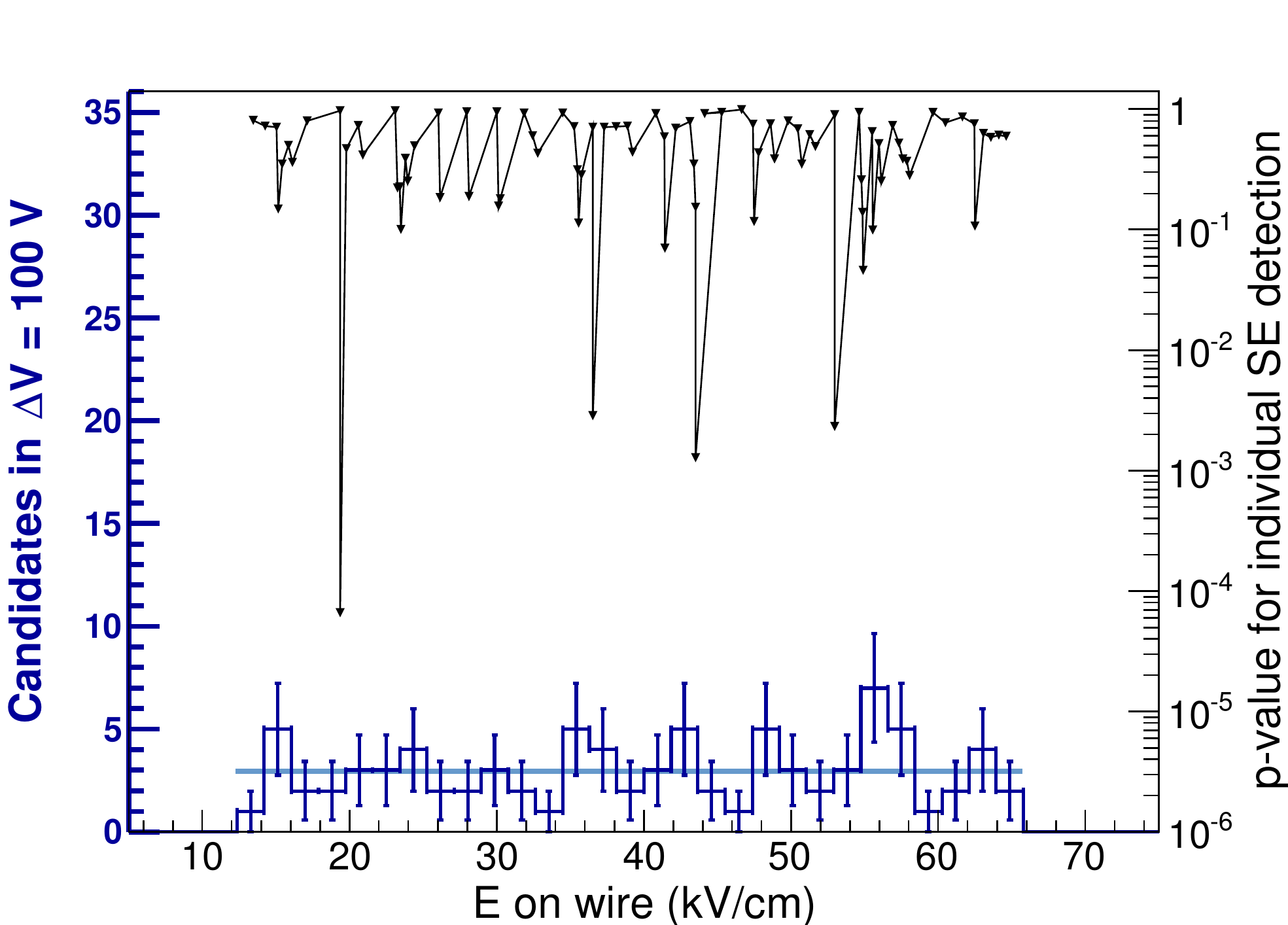}
\includegraphics[width=8.6cm]{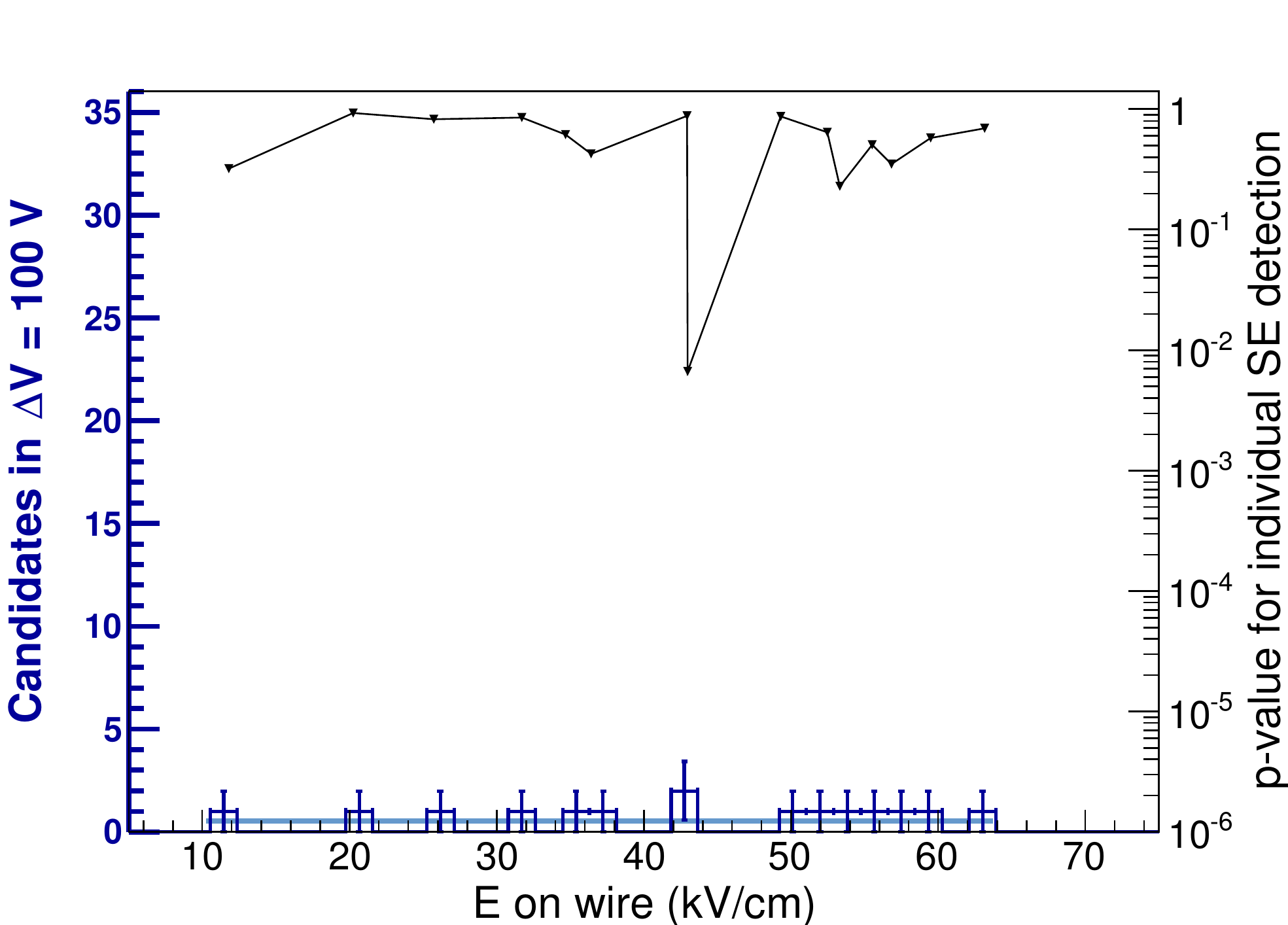}
\end{center}
\caption{First voltage ramping test on treated 200-$\mu$m LUX cathode wires: Left -- Electropolished (Run~19); Right -- Acid-cleaned (Run~25). Compare with untreated samples in Figs.~\ref{fig:Run15_pvalues} and~\ref{fig:Run16}.
}
\label{fig:treatmets}
\end{figure*}

The electropolished SS302 LUX cathode sample did not reveal extended emitters during any voltage ramping test---see example in Fig.~\ref{fig:treatmets} (left). A modest improvement in faint emission and fewer impulsive emitters were also observed. An electropolished sample of the ZEPLIN-III SS316L wire showed a more marked improvement in faint emission (cf.~Run~20 in Table~\ref{tab:summarytable}). In conclusion, the electropolished SS302 sample became more resilient, out-performing any untreated wire tested with high sensitivity, but the electropolished SS316L wire improved even more markedly.

\subsubsection{Acid cleaning and passivation}

These are two distinct, yet complementary processes for the finishing of metal surfaces. The removal of oxides and surface corrosion products by the action of an acid solvent is commonly known as `pickling'. When nitric acid is employed and the bath is prolonged and enhanced by temperature and some form of cavitation, the process is termed `decontamination' or simply `acid cleaning'. Such a bath not only removes impurities from the stainless steel, but it also dissolves iron atoms from the surface, thus decreasing its propensity for corrosion. The acid cleaning will then promote a chromium-enriched surface, more stable chemically and potentially more homogeneous. The oxidation of the new surface, i.e.~the `passivation', will unavoidably follow any pickling. The oxidation must be completed quickly and in a clean environment if an impurity-free, smooth chromium oxide layer is to be achieved.

Two samples were tested after ultrasonic cleaning in a 35\%~HNO$_3$ bath at $45^\circ\textmd{C}$ for 30~minutes. The samples were then immediately immersed in deionized water for passivation, dried with a nitrogen gun, and sealed in an air atmosphere. The whole process was conducted in a Class 1000 cleanroom; the wires had been previously cleaned according to the usual procedure.

This treatment was remarkably successful in eliminating electron emission, as shown in Fig.~\ref{fig:treatmets} (right) for Run~25. Over four runs conducted on passivated samples of the LUX cathode wire, no extended emitters were observed and only one double-candidate waveform was seen (at 53~kV/cm, during the third run of one sample). In addition, the major decrease in faint emission confirmed that this phenomenon actually originates in the wire itself, rather than being a background process in our measurement. To confirm that the chamber was still operating nominally, this first test of an acid-cleaned wire was followed by Run~26 using an untreated sample, which exhibited a typical emission pattern.  

Although both acid-treated samples were produced at the same time, they have different histories. The first was installed in the chamber only two days after the treatment. The second was tested after ageing in (bagged) air for 8~months, and then kept in vacuum between runs thereafter. The small number of samples and the scant evidence for emission make the following observations somewhat uncertain, but the data seem to suggest that: i) the wire performance was slightly better for the freshly-treated sample; ii) for the aged sample it was better in the first run than in the last one; and iii) faint emission increased very slowly with field. Despite the major improvement in performance with respect to any other (non-treated) sample, these observations hint at the possibility of some modest deterioration associated with ageing and electrical stress. Note that the wires are stressed three times per run and to higher fields than would normally be experienced in the LZ detector.

\subsubsection{Gold-plating} \label{subsec:gold-plating}

Gold-plated tungsten wire has traditionally been used in wire chambers, with its expected resilience attributed to the high work function and resistance to corrosion of the gold layer. We conclude this section with the test of one such wire with 125$~\mu$m diameter. Unfortunately, this spool was decades old and of uncertain provenance. The performance during the first ramping test was indeed found to be better than that with any untreated stainless steel wire, and better than that of the electropolished samples too---although not as good as the acid-cleaned samples. However, this performance deteriorated somewhat in the second and third ramping tests, contrary to the trend seen with the untreated samples in our study. Whereas for the first test only two impulsive emitters were registered, four were observed during the second test and at lower fields ($<$60~kV/cm). In the third test, two extended emitters were detected at 94 and 99~kV/cm, despite the lower sensitivity settings. These observations suggest a clear benefit from the gold-plating, but some lack of resilience against electrical stress.

%%%%%%%%%%%%%%%%%%%%%%%%%%%%%%%%%%%%%%%%%%%%%%%%%%%%%%%%%%%%%%%%%%
\section{Discussion}

Clear electron emission has been observed from most wires tested in this study, and our results are broadly consistent with the highest surface fields which several LXe-TPCs were able to sustain using precisely the same wires. These experiments operated close to the onset of significant emission, at surface fields in the range 40--65~kV/cm~\footnote{We estimate the average electrostatic fields on the wire surfaces using the calculations for parallel wire grids in Refs.~\cite{Dahl,McDonald}, with grid parameters and operating voltages as published by those experiments.}:  ZEPLIN-III and its prototypes~\cite{Akimov02,Howard04,Akimov07,Burenkov09,Lebedenko09,Akimov12b} (all using SS316, 100~$\mu$m diameter); and the LUX gate electrode~\cite{LUXr2,Akerib14} (SS304, 100~$\mu$m). Approximately in this range is where we observe clear extended emitters from the same wires in our study. Detailed measurements of electron emission from the LUX gate were reported in Ref.~\cite{AdamThesis}, which shows very similar emission phenomena to those reported here for well-resolved emitters localised at various positions on the grid, starting at fields of $\sim$60~kV/cm. The LUX cathode (SS302, 200~$\mu$m) performed less well, sustaining only 19~kV/cm in LUX Run~3~\cite{Akerib14}, and this is also the most prolific emitter identified in our study, with extended emitters identified at fields as low as 10~kV/cm. Finally, we confirmed that the Xed detector cathode wire (BeCu, 40~$\mu$m), which operated at 220~kV/cm in the actual experiment~\cite{Shutt07}, was able to reach at least 310~kV/cm in our earlier tests (albeit with limited sensitivity) and sustained 405~kV/cm without photon emission, suggesting good electrical resilience. We point out that this phenomenology extends to experiments featuring etched-mesh electrodes, including XENON100~\cite{XENON100instrument} (surface fields in that experiment are also estimated in Ref.~\cite{AdamThesis}).

Note that all LXe-TPCs referenced above had SE sensitivity and so these are appropriate comparisons, both between them and with our tests. Difficulties with HV delivery are also well documented in other LAr- and LXe-TPCs for dark matter, double-beta decay and neutrino detection, but onset fields cannot be accurately compared as they did not operate with quantum-level sensitivity.

Away from the main emitters our measurements reveal a variable level of faint emission in all untreated samples. We are confident that this excess rate is not due to background in our measurements as the average emission rates are very low indeed with some treated wires. The residual SE rate was measured in low-background conditions in both runs of the ZEPLIN-III experiment~\cite{Santos11}; in this 3-electrode LXe-TPC there was a single cathodic grid using 117~m of wire. The electron rate was found to be as low as $\sim 1$~c/s for a wire field of $\sim 38$~kV/cm in the lowest background configuration of the experiment. This is clearly incompatible with the faint emission rate of $\approx$30~c/s for the 52~mm length of wire under study. Thus, we conclude that much of this faint emission is transient in nature. In contrast to the emitters discussed above, this phenomenon would not have prevented the operation of these detectors, but it will have affected their sensitivity for `S2-only' physics searches conducted by electron counting near the ionisation detection threshold. Understanding and mitigating these residual emission rates were key objectives of this study.

Another main result from our study is the confirmation that no {\em intrinsic} threshold exists leading to electrical breakdown of cathodic surfaces in LXe at fields at least as high as 160~kV/cm, with weaker constraints at 405~kV/cm from one of the studied samples. On the other hand, a lower emission threshold could not be identified either, especially as most of our tests were restricted to fields $\gtrsim$10~kV/cm. We see some signs of emission at the lowest fields in practically every case. 

\vspace{3mm}
From the plethora of published data and from our own results, the macroscopic field on the wire surface does not appear to be the key parameter that explains most phenomenology. The field dependence of all types of emitter is actually quite modest. For example, the 200~$\mu$m LUX cathode wire did not perform much better in our tests than the 100~$\mu$m wire used in the LUX gate and in the cathode plane installed initially, despite the lower fields on the former; this agrees with the actual experience reported by LUX~\cite{LUXr2}. On the other hand, there was variability for untreated samples of the same material. This could be explained by the very small surface areas tested in our study. The concept of \textit{stressed area}~\cite{Lockwitz14} combines surface fields with the area of the cathodic surface (typically applied to large electrodes); the 200~$\mu$m wire holds a lower field over a larger area, but the stressed area is comparable in both cases. This idea is therefore consistent with our results in the limit of small areas and low statistics. It will certainly be important to study grids made with much longer wires; the lengths used in detectors such as LZ are thousands of times longer than those under scrutiny here. LZ is conducting such a programme~\cite{LZTDR}.

It has been proposed that the Malter effect~\cite{Malter36} could be one reason behind the HV breakdown in these detectors. This effect explains some electrical breakdown phenomena in classical wire chambers operating at high luminosity through the accumulation of positive ions on the oxide layers coating the cathodic wires. However, the ion rates in LXe-TPCs used for rare event searches are very low, and the spurious emission observed in these detectors is not that dissimilar when operating on the surface or in very low-background conditions in underground laboratories---or, for that matter, during calibration at higher rates of energy deposition. In addition, the Malter effect involves relatively long time constants for charging and discharging of the oxidised cathodic surfaces. This is contrary to the behaviour of our typical emitters, which exhibit very impulsive outbreaks and equally fast decays. As much as understanding the dynamical behaviour of positive ions in these detectors is an important topic, we do not believe that this explains the main effects reported in this study.

The abruptness of the emission onset is suggestive of a quantum mechanical tunnelling effect. The unpredictability of each onset points towards a localised, microscopic origin. In fact, the marked improvement with surface treatment suggests that the causes behind these microscopic emitters are associated with some type of surface defect. Combining an area scaling with some underlying distribution of defect parameters could lead to a macroscopic description which is consistent with the stressed area concept, or even with the perceived appearance of an onset near some specific field for particular geometries and materials. Indeed, it has been widely believed that this phenomenology could be explained by field emission from very small metallic protrusions or filaments on the wire surface, following the identification in older studies of very steep Fowler-Nordheim I-V curves~\cite{FN28} for currents in the range $10^{-10}$--$10^{-13}$~A (cf.~\cite{Halpern69}; see \cite{Noer82} for a review). We were not able to derive accurate I-V curves from our measurements and our results cannot be directly compared with the classical studies on field emission at much higher currents. While our fields are not sufficient for a measurable field emission from a perfect surface with work function as high as that of stainless steel (4.3~eV)---even considering a 0.61~eV reduction due to the electron affinity of LXe~\cite{Tauchert75}---the introduction of protrusions, each characterised by a `field enhancement factor' $\beta$ (the ratio of the locally enhanced field to the average field for the perfect geometry), is an attractive explanation that could conceivably lead to the minute currents we observe ($\sim$$10^{-16}$~A). Enhancement factors $\beta\!\sim$100--1000 can be estimated for extremely long and sharp (nanometre-sized) filaments~\cite{Alpert64,Vibrans64}, and such large factors would be required to explain macroscopic emission currents from stainless steel wires in LXe. Our SEM micrographs cannot rule out features of this size over significant wire lengths, and so we cannot exclude that such small filaments might be present in our samples (although that is unlikely given the high-quality finish of TPC wires). In specially-designed experiments where the microscopic emitter could be imaged, non-protruding defects (such as cracks, scratches, `micro-craters', grain boundaries or impurity inclusions) are often found instead, whereas protrusions are more likely to be a consequence of previous sparks occurring during the test~\cite{Latham70,Biradar70,Beukema81}---our samples were not exposed to such discharging.

There are key pieces of evidence against such explanations relying on small protruding filaments. Firstly, all emitters disappear whilst sustaining currents far lower than those necessary to burn off these filaments, no matter how small. Secondly, the enhanced local fields at viable emission sites (i.e.~those predicted to generate measurable tunnelling currents) would easily exceed both the light and the charge multiplication thresholds in LXe, and therefore a profusion of photons would be expected, along with multi-electron signals in all cases; we observe only tenuous evidence for photon emission in one case, and our emitters are consistent with single electrons. Finally, one would expect electropolishing to be particularly effective in attacking these filaments but the improvement after this treatment is reasonably modest. Therefore, we do not believe that the presence of filamentary protrusions on the wires provides a reasonable explanation.

The issue of the electric field strength at the emission site is worth exploring in more detail. Studies during the LUX grid conditioning campaign~\cite{AdamThesis} detected the presence of locally-enhanced fields for well-localised emitters on the gate grid. These emitters were similar in onset field and in intensity to those observed in this work, and they were seen to persist for minutes or even hours at constant field. Some of the waveforms recorded in that study contained multiple electrons per pulse emitted from the same point on the gate grid. The multiplicity distributions were well described by a Polya function, which is characteristic of charge avalanche phenomena. Most emitters showed signs of modest charge multiplication, which in the liquid phase has a threshold of $\sim$700~kV/cm. Such an effect requires only $\beta\!\sim$10--20. In addition, evidence for simultaneous photon production was also consistently detected, with the number of photons emitted being comparable with the number of initial electrons, in reasonable agreement with the liquid-phase electroluminescence expected at those fields. Similar evidence for enhanced fields is also seen in one of our emitters, from which photon production was detected simultaneously with electron emission (although we saw no evidence of charge multiplication). In conclusion, although there is evidence of $\beta$ factors of $\sim$10--20, these are not sufficient to lead to significant rates of field emission from stainless steel.

Perhaps the most significant result from this study was the demonstration that the wire surface quality---and in particular its resistance to corrosion---is the main factor determining their electrical resilience. Amongst the various types of stainless steel, the most resistant to corrosion is SS316L and this systematically out-performed the other types in our study (SS304 and SS302). The improvement after chemically treating the SS302 wire was striking, and it had persisted several months after treatment, suggesting that the uniform chromium oxide layer promoted by the acid cleaning is long-lasting and strong enough to be of practical interest for the noble liquid TPCs. Electropolishing also improved the wire performance, but not as dramatically; note that the electropolishing procedure also involved a fast pickling and passivation stage. All of these results suggest that corrosion products---possibly the ferrous contaminants---could be responsible for the degradation of electrical performance in all cases. The microscopic phenomena underlying this effect are less clear; they could be related to surface imperfections or domain boundaries between the ferric and chromic oxides in the untreated wire, or to other causes. In some cases the oxide layer---and its irregularities---have been seen to facilitate emission~\cite{Yang91}. Oxide monolayers can lead to a resonant tunnelling effect~\cite{Duke67,Gadzuk70}, and this is an appealing explanation for our observations.

Other materials and coatings less prone to corrosion were also shown to achieve good performance; in particular, the BeCu wire performed well. Results for the gold-plated tungsten were less conclusive, but we tested only one sample of this wire. Note that there may be reasons why stainless steel wires could still be preferred to materials such as BeCu, e.g.~mechanical properties (tensile strength, thermal contraction) or neutron production rates via $(\alpha,n)$ reactions on beryllium.

Finally, we discuss briefly the issue of conditioning, whereby the number and intensity of emission centres decreases after the sample is subjected to periods of sustained electrical stress or controlled discharging. This is a common technique for the `HV training' of the classical wire chambers; it can also signify time-dependent effects at constant field. In typical field emission experiments the interruption or modification of the I-V curve results from micro-sparks involving peak currents of at least $10^{-9}$~A~\cite{Halpern69,Noer82}; during such a conditioning process, the `burning off' of emitters is often mentioned. However, we note that in LXe-TPCs the cathodic electrodes are immersed in the liquid phase, and the currents registered in our study are very far indeed from being able to dissipate the power required to vaporise or damage even nanometre-sized emission centres. Nevertheless, a mild conditioning effect was observed when comparing the initial ramping test with subsequent ones for our wire samples; some modest success was also achieved in the LUX detector which allowed it to operate at increased gate voltage~\cite{Akerib14}. Significantly, it is also the case that extrapolating the distribution of emitter onsets recorded in the very short wires in our tests to the lengths required in larger detectors suggests that some form of condition is indeed at play. The term `conditioning' may cover a collection of distinct effects, acting on different current regimes and on different types of emission centre. We should certainly distinguish between gas-phase conditioning and the effect observed in our study, where the wire is in the dense liquid: even in the absence of macroscopic discharging, in the gas phase there is the possibility for atom sputtering from ion back-flow (which becomes significant at low pressures, leading to both higher currents from charge multiplication and to higher ion energies on impact). In addition, we cannot rule out that some of these improvements could be temporary, possibly related to the depopulation of surface electronic states.

%%%%%%%%%%%%%%%%%%%%%%%%%%%%%%%%%%%%%%%%%%%%%%%%%%%%%%%%%%%%%%%%%%
\section{Conclusion}

Spurious electron emission from thin cathodic wires immersed in LXe has been studied at electric fields ranging from 10~kV/cm to 163~kV/cm in the current regime $10^{-18}$--$10^{-15}$~A using single electron counting. We tested mostly wires with well understood provenance and which had been used in several LXe-TPCs with similar levels of sensitivity, all operating not far from instability.

No macroscopic discharging occurred up to 405~kV/cm and no evidence was observed for an intrinsic threshold specific to the metal-LXe interface. Three types of emission were registered in untreated wires: well-defined emitters extending over a narrow, but measurable range of field; more impulsive ones registering very high instantaneous rates which appear and disappear over a short time; and a more pervasive rate of faint emission present at all fields. This phenomenology appears to be consistent with the effects which have limited the electrical performance of noble liquid TPCs for well over a decade.

The comparison of different wire materials, in particular the stainless steel varieties, suggests that the electrical resilience is correlated with the resistance to corrosion. A major result of this study was that all emitter types were significantly suppressed in stainless steel samples extensively cleaned with nitric acid and passivated, bringing about at least order-of-magnitude improvements in overall electron emission rates. This suggests that corrosion products on the oxide layer could be the leading cause of the spurious emission, which is radically decreased when these are replaced by the high-quality chromium oxide layer promoted by the acid cleaning.

This and other studies rule out some microscopic causes for at least some types emission, including field emission from very small filamentary protrusions and the Malter effect related to Xe$^+$ ion accumulation on the cathode. There is, however, evidence for enhanced fields at some emission sites, as we observe simultaneous photon and electron emission in one instance (a previous study \cite{AdamThesis} documented photon emission more extensively and found also limited charge multiplication in some cases; both phenomena have thresholds of many hundreds of kV/cm). Further work is required to understand the precise microscopic emission mechanism, its temporal dependence (including the effect of conditioning) and its scaling with wire length. Additional work to explore even more effective and durable wire treatments and coatings would also be beneficial. 

\vspace{5mm} %5mm vertical space
In conclusion, this study was able to progress significantly our understanding of this complex phenomenology, provide some guidance as to the best performing wire materials, and it offers an effective and practical way to treat stainless steel wire grids to improve the HV performance of LXe detectors such as LZ. In addition, the mitigation of the faint emission observed at all fields is likely to decrease the spurious electron background arising in the cathodic electrodes and improve our ability to conduct rare-event searches by single electron counting at very low energies ($<$1~keV).

%%%%%%%%%%%%%%%%%%%%%%%%%%%%%%%%%%%%%%%%%%%%%%%%%%%%%%%%%%%%%%%%%%
\section*{Acknowledgements}
We wish to thank our collaborators in the LZ dark matter experiment for their ongoing support---and particularly T.~Shutt and J.~Va’vra at SLAC, W.~Waldron at LBNL, D.~McKinsey, E.~Bernard and L.~Tvrznikova at U.C. Berkeley, R.~Mannino from TAMU and P.~Majewski from RAL for useful discussions and for providing various wire samples. Thanks are due also to V.~Chepel from LIP-Coimbra for useful comments on this manuscript. We thank S.~Dardin at LBNL for help with SEM imaging and electropolishing of wire samples, and to J.~Dobson at UCL for assistance with acid cleaning. Thanks are due to F.~Neves and V.~Solovov from LIP-Coimbra for help with the ZEPLIN-III slow control electronics. We acknowledge contributions from Imperial students S.~Lennon, C.~Laner and W.~Sio. We thank D.~Williams and colleagues at the Physics Mechanical Instrumentation Workshop and D.~Clark of the HEP group workshop at Imperial for hardware manufacture. We thank The Welding Institute for their assistance with e-beam welding of some components. This work was supported by the U.K. Science \& Technology Facilities Council (STFC) under award numbers ST/K006428/1 and ST/M003655/1, and by PhD studentship ST/K502042/1 (AB).

%%%%%%%%%%%%%%%%%%%%%%%%%%%%%%%%%%%%%%%%%%%%%%%%%%%%%%%%%%%%%%%%%%
\vspace{8mm}
\section*{References}
\bibliography{mybibfile}

\end{document}